\documentclass[preprint,1p]{elsarticle}

\usepackage{amsmath}
\usepackage{amssymb}
\DeclareMathOperator{\sgn}{sgn}
\usepackage[braket]{qcircuit}
\usepackage{algorithm}
\usepackage{listings} 
\usepackage{algpseudocode}
\usepackage{siunitx}
\usepackage{booktabs}
\usepackage{tabularx}
\usepackage{amsthm}
\usepackage{CJK}
\usepackage[unicode]{hyperref}
\hypersetup{
    colorlinks=true,
    linkcolor=blue,
    urlcolor=blue
}
\usepackage{enumerate}
\usepackage{subfig}
\usepackage{tikz}

\newdefinition{rmk}{Remark}
\newdefinition{problem}{Problem}
\newtheorem{thm}{Theorem}
\newproof{pf}{Proof}

\journal{Computer Methods in Applied Mechanics and Engineering}

\begin{document}

\begin{frontmatter}
\title{Enabling Large-Scale and High-Precision Fluid Simulations on Near-Term Quantum Computers}

\cortext[contr]{These authors contribute equally to this work.}
\cortext[cor]{Corresponding author}

\author[1]{Zhao-Yun~Chen\corref{contr}}
\author[2]{Teng-Yang~Ma\corref{contr}}
\author[2]{Chuang-Chao~Ye\corref{contr}}
\author[3]{Liang~Xu}
\author[4]{Wen~Bai}
\author[5]{Lei~Zhou}

\author[6]{Ming-Yang~Tan}

\author[7]{Xi-Ning~Zhuang}
\author[7]{Xiao-Fan~Xu}
\author[7]{Yun-Jie~Wang}
\author[7]{Tai-Ping~Sun}
\author[7]{Yong~Chen}
\author[7]{Lei~Du}
\author[7]{Liang-Liang~Guo}
\author[7]{Hai-Feng~Zhang}
\author[7]{Hao-Ran~Tao}
\author[7]{Tian-Le~Wang}
\author[7]{Xiao-Yan~Yang}
\author[7]{Ze-An~Zhao}
\author[7]{Peng~Wang}
\author[7]{Sheng~Zhang}
\author[7]{Ren-Ze~Zhao}

\author[2]{Chi~Zhang}
\author[2]{Zhi-Long~Jia}
\author[2]{Wei-Cheng~Kong}
\author[2]{Meng-Han~Dou}

\author[8]{Jun-Chao Wang}
\author[7]{Huan-Yu~Liu}
\author[1]{Cheng~Xue}
\author[6]{Peng-Jun-Yi~Zhang}
\author[Huang]{Sheng-Hong~Huang}

\author[7]{Peng~Duan}
\author[7]{Yu-Chun~Wu}
\author[7]{Guo-Ping~Guo\corref{cor}}
\ead{gpguo@ustc.edu.cn}

\affiliation[1]{organization = {Institute of Artificial Intelligence, Hefei Comprehensive National Science Center},
                city = {Hefei},
                postcode = {230088},
                country = {China}}

\affiliation[2]{organization = {Origin Quantum Computing Technology (Hefei) Co., Ltd.},
                city = {Hefei},
                postcode = {230088},
                country = {China}}
                
\affiliation[3]{organization = {China Academy of Aerospace Aerodynamics},
                city = {Beijing},
                postcode = {100074},
                country = {China}}

\affiliation[4]{organization = {Chinese Aeronautical Establishment},
                city = {Beijing},
                postcode = {100012},
                country = {China}}

\affiliation[5]{organization = {Xi'an Aeronautics Computing Technique Research Institute, AVIC},
                city = {Xi'an},
                postcode = {710000},
                country = {China}}

\affiliation[6]{organization = {Department of Modern Mechanics, University of Science and Technology of China},
                city = {Hefei},
                postcode = {230026},
                country = {China}}

\affiliation[7]{organization = {CAS Key Laboratory of Quantum Information, University of Science and Technology of China},
                city = {Hefei},
                postcode = {230026},
                country = {China}}

\affiliation[8]{organization = {Laboratory for Advanced Computing and Intelligence Engineering},
                city = {Zhengzhou},
                postcode = {450000},
                country = {China}}

\affiliation[Huang]{organization = {CAS Key Laboratory of Mechanical Behavior and Design of Materials, University of Science and Technology of China},
                city = {Hefei},
                postcode = {230026},
                country = {China}}

\begin{abstract}
Quantum computational fluid dynamics (QCFD) offers a promising alternative to classical computational fluid dynamics (CFD) by leveraging quantum algorithms for higher efficiency. This paper introduces a comprehensive QCFD method, including an iterative method ``Iterative-QLS'' that suppresses error in quantum linear solver, and a subspace method to scale the solution to a larger size. We implement our method on a superconducting quantum computer, demonstrating successful simulations of steady Poiseuille flow and unsteady acoustic wave propagation. The Poiseuille flow simulation achieved a relative error of less than $0.2\%$, and the unsteady acoustic wave simulation solved a 5043-dimensional matrix. We emphasize the utilization of the quantum-classical hybrid approach in applications of near-term quantum computers. By adapting to quantum hardware constraints and offering scalable solutions for large-scale CFD problems, our method paves the way for practical applications of near-term quantum computers in computational science.
\end{abstract}

\begin{keyword}
Quantum computational fluid dynamics\sep
Superconducting quantum computer\sep
Variational quantum linear solver\sep
Noisy intermediate-scale quantum
\end{keyword}
\end{frontmatter}

\section{Introduction}\label{sec:1}
Computational fluid dynamics (CFD) is crucial for flow mechanism research and industrial design. Advances in computing have continually introduced new methodologies and paradigms in scientific computation. Initially, CFD methods based on the finite volume and finite difference methods were developed for CPU-based high-performance computers. Over the past decade, GPU-based heterogeneous computing has gained prominence, enabling complex and computationally expensive methods like the spectral element method~\cite{xu2022spetral}, high-order finite volume/difference methods~\cite{fu2016teno,Ji2022}, and machine learning-based methods~\cite{mao2020pinn,XIONG2023105811}. However, simulating extreme flow parameters and large-scale models remains unaffordable with the most advanced classical supercomputers, which are hitting performance bottlenecks due to the atomic scale of transistors. This has motivated the exploration of new computing paradigms, with quantum computing being one of the most promising. 

Significant progress in quantum computing has encouraged innovative methods in CFD~\cite{succi2023quantum}. Currently, there are two major approaches to performing fluid simulations on quantum computers. One methodology emulates the flow field by a quantum state, simulating the fluid dynamics via Hamiltonian simulation using the hydrodynamic Schrödinger equation, as shown in~\cite{meng2023quantum, meng2024simulating, PhysRevA.105.052404}. Alternatively, a broader approach replaces specific classical algorithms in CFD with their quantum counterparts to tackle sub-problems~\cite{cao2013quantum, childs2020quantum}. Of particular interest are linear systems of equations, which are central to many classical CFD methods~\cite{hesthaven2007spectral,cockburn2003,Butcher1976} and often the most time-consuming and resource-intensive components, thus constraining large-scale fluid simulations. Quantum linear solvers (QLS), as demonstrated in Refs.~\cite{Harrow2009,An2022,Costa2022}, provide an exponential speedup over classical methods, making them a potential solution for performing fluid simulations on quantum computers. 

Concurrently, advancements in quantum hardware have demonstrated advantages in specific problems experimentally~\cite{arute2019quantum,madsen2022quantum,Ren_2022}. With current quantum computers now having several hundred qubits, near-term quantum computing applications for challenging flow problems are foreseeable. Nevertheless, quantum computational fluid dynamics (QCFD) still faces significant challenges when implemented on near-term quantum devices, specifically noisy intermediate-scale quantum (NISQ) devices~\cite{preskill2018quantum}. A major hurdle stems from quantum errors. Most quantum algorithms are conceived within the context of the fault-tolerant quantum computation (FTQC) era~\cite{preskill1998reliable, campbell2017roads}, wherein noisy physical qubits are redundantly encoded to form an error-free ``logical qubit"~\cite{devitt2013quantum}. However, only a few logical qubits have been successfully demonstrated to date~\cite{sivak2023real, ni2023beating}, implying that quantum errors will remain a significant concern for near-term quantum computers. Error suppression methods, such as quantum error mitigation (QEM)~\cite{cai2023quantum}, are therefore crucial for practical quantum applications~\cite{kim2023evidence}. In the context of QCFD, the temporal evolution of flows means that computational errors can accumulate and amplify, potentially obscuring the true flow dynamics. Hence, ensuring computational accuracy is paramount. 

Despite these challenges, several noteworthy contributions have been made to achieve fluid simulations on NISQ devices. One idea is to achieve a high-quality solution by utilizing high-fidelity qubits, demonstrated by two-dimensional unsteady flows discretized spatially with 1024 grid points performed on a well-calibrated superconducting quantum computer, achieving the largest quantum fluid simulation reported so far~\cite{meng2024simulating}. Another idea is to find error-resilient quantum methods, introducing the variational framework to solve linear systems on near-term quantum computers, such as variational quantum linear solvers (VQLS)~\cite{cerezo2021variational, XU2021, Bravo_2023}. This approach has been applied to address problems such as potential flows, Stokes flows, lid-driven cavities, and the advection-diffusion equation~\cite{jaksch2023variational, song2024incompressible, liu2024variational, demirdjian2022variational, pool2024nonlinear, PhysRevA.101.010301}. These works successfully demonstrate the feasibility of QCFD on NISQ devices. However, current methods still face scalability issues on near-term quantum computers, where quantum errors still hinder the current method from computing a larger-scale fluid flow. Therefore, a more effective way to suppress the quantum error should be proposed to enable large-scale quantum fluid simulation. 

In this paper, we propose a comprehensive method that enables large-scale and high-precision fluid simulation on near-term quantum computers. To demonstrate the ability of the method, the steady Poiseuille flow and unsteady acoustic wave propagation are simulated on a superconducting quantum computer. This algorithm can be regarded as a bridge between quantum and classical computers. According to our findings, our method can adapt to the quantum computer's capabilities, enabling collaboration between quantum and classical computers. Although the current solution uses only a limited number of qubits due to hardware constraints, advancements in quantum computing offer the possibility of investing more quantum resources to achieve quantum speedup. Our approach is also a general-purpose quantum-enhanced linear solver method, applicable to other potential applications in computational science, including but not limited to finite element analysis, computer graphics, and signal processing. With successful, high-precision demonstrations of large-scale CFD problems on a real quantum computer, this paper marks a step towards scalable quantum-enhanced solutions for CFD and paves the way for the practical application of near-term quantum computers.

The remaining part is organized as follows. In Sec.~\ref{sec2}, basic concepts in quantum computing and the technical path of QCFD using QLS, as well as challenges of implementing fluid simulation on near-term devices, are introduced. Sec.~\ref{sec3} presents the details of our proposed method. Flow simulation experiments on a real quantum computer are shown in Sec.~\ref{sec:experiments}. Concluding remarks are given in the final section. 

\section{Preliminaries} \label{sec2}
\subsection{Basic concepts in quantum computing}

\paragraph{Quantum information}
The basic information unit in quantum information is the qubit, which is analogous to a classical bit but with the ability to exist in a superposition of states. Mathematically, a single qubit state can be represented as a vector in two-dimensional Hilbert space:
\begin{equation}
\left| \psi \right\rangle = \alpha \left| 0 \right\rangle + \beta \left| 1 \right\rangle,
\end{equation}
where \(\left| 0 \right\rangle\) and \(\left| 1 \right\rangle\) are the basis states, and \(\alpha\) and \(\beta\) are complex numbers satisfying \(|\alpha|^2 + |\beta|^2 = 1\).
The state of a multi-qubit system is described by the tensor product of individual qubit space, which means an $n$-qubit quantum state represents a vector in $2^n$-dimensional Hilbert space. 

\paragraph{Quantum computing}

In quantum computing, we perform various operations on qubits, which store quantum information, to accomplish specific computational tasks. Quantum computing has different computational models, such as quantum annealing, one-way quantum computing, and quantum circuits, each of which can theoretically be converted into the others~\cite{nielsen2010quantum}. Quantum circuit model is one of the most common models of quantum computing. The basic unit of a quantum circuit is the quantum logic gate, with each gate representing a unitary matrix. Common quantum logic gates include single-qubit gates and multi-qubit gates. Single-qubit gates include the X gate and the Hadamard gate, with their specific matrix forms as follows:
\begin{equation}
X = \begin{pmatrix}
0 & 1 \\
1 & 0
\end{pmatrix}, \quad
H = \frac{1}{\sqrt{2}} \begin{pmatrix}
1 & 1 \\
1 & -1
\end{pmatrix}.
\end{equation}
Two-qubit gates include the CNOT gate and the CZ gate, with their matrix forms as follows:
\begin{equation}
\text{CNOT} = \begin{pmatrix}
1 & 0 & 0 & 0 \\
0 & 1 & 0 & 0 \\
0 & 0 & 0 & 1 \\
0 & 0 & 1 & 0
\end{pmatrix}, \quad
\text{CZ} = \begin{pmatrix}
1 & 0 & 0 & 0 \\
0 & 1 & 0 & 0 \\
0 & 0 & 1 & 0 \\
0 & 0 & 0 & -1
\end{pmatrix}.
\end{equation}

A quantum logic gate can process a \(2^n\)-dimensional quantum state with one step, achieving coherent superposition and interference between each dimension of the target quantum state. These features of quantum logic gates can be utilized to construct quantum algorithms that accelerate the solution of specific problems. Numerous quantum algorithms have been proposed to accelerate the solution of various problems, with applications spanning mathematics, physics, chemistry, biology, computer science, and more. Typical problems include integer factorization, search, Hamiltonian simulation, matrix inversion, classification, prediction, and others. For more information, refer to the \href{https://quantumalgorithmzoo.org}{Quantum Algorithm Zoo}.

\subsection{The technical path of QCFD using QLS} \label{sec:1-2}

Solving linear equations is a fundamental step in the study of CFD problems. A major branch of QCFD aims to accelerate this process by QLS. This process involves several key steps: discretization, linearization, and the application of QLS.

In CFD, one must solve linear or nonlinear partial differential equations, such as the Navier-Stokes equations. These equations can be represented as:
\begin{equation}
    \partial{\boldsymbol{Q}}/\partial{t}=\mathcal{F}(\boldsymbol{Q}),
\end{equation}
where $\boldsymbol{Q}$ is the variable vector, and $\mathcal{F}(\boldsymbol{Q})$ represents linear or nonlinear operators. The temporal and spatial derivatives are discretized with explicit or implicit schemes. While explicit methods have strict stability requirements limiting the time step, implicit methods are preferred for their ability to handle larger time steps and provide numerical stability, particularly for stiff differential equations. Therefore, we use implicit methods, which are formulated as:
\begin{equation}\label{eq-time-discretize}
    \frac{\boldsymbol{Q}^{t+1}-\boldsymbol{Q}^t}{\Delta t} = R(\boldsymbol{Q}^{t+1}),
\end{equation}
where $\Delta t$ represents the time step, and $R(\boldsymbol{Q}^{t+1})$ is the residual. Generally, when $\mathcal{F}(\boldsymbol{Q})$ is nonlinear, this leads to a nonlinear system of equations for $\boldsymbol{Q}^{t+1}$, necessitating a linearization procedure.

Linearization is crucial for approximating these nonlinear equations into a system of linear equations, which can be global or local. Global linearization techniques, such as Carleman linearization~\cite{kowalski1991nonlinear,liu2021efficient}, coherent state linearization~\cite{kowalski1994methods,engel2021linear}, and Koopman-von Neumann linearization~\cite{koopman1931hamiltonian,joseph2020koopman}, approximate the entire system into a finite-dimensional linear system. On the other hand, local linearization methods, like the Newton-Raphson method~\cite{hildebrand1987introduction}, transform the nonlinear equation into an iterative process of solving linear systems.

The final step involves solving these linear equations using QLS. The selection of an appropriate QLS depends on the specific CFD problem, considering the problem's scale and the quality of the quantum chip. By executing the QLS in conjunction with classical computations, we complete the solution of the CFD problem, leveraging the power of quantum computing to enhance computational efficiency.

\subsection{Challenges of implementing fluid simulation on near-term devices}\label{sec:3}

Currently, several QCFD algorithms have been proposed, but there are still challenges when solving specific CFD problems on near-term quantum computers, i.e. NISQ devices. Here, we list and analyze these challenges.

Firstly, due to the impact of quantum noise, the quantum state produced by a quantum circuit executed on NISQ devices is prone to errors. This inaccuracy results in low-precision, unreliable solutions and subsequently hinders the computation of unsteady flow using time-stepping methods.

Secondly, the number of qubits on NISQ devices is limited. This limitation arises not only because the number of qubits that can be fabricated on a quantum processor is restricted, but also because quantum noise limits the size of the circuit, preventing the number of qubits used in a quantum circuit from increasing~\cite{cross2019validating}.

Thirdly, due to the ``no-cloning theorem''~\cite{nielsen2010quantum}, transferring quantum states to classical computers is challenging because the quantum state collapses after each measurement. Extracting the quantum state to classical data requires rerunning the entire computation for each subsequent measurement. This process, known as ``quantum tomography'', is time-consuming and presents a significant obstacle to realizing practical quantum applications~\cite{nielsen2010quantum}. Improving the efficiency and precision of quantum tomography methods is crucial for achieving quantum advantage. $l_2$-norm tomography~\cite{kerenidis2020quantum} reconstructs quantum states by minimizing the average error between the measured data and the model prediction but is sensitive to large outliers, which can distort the reconstruction if not properly managed. In contrast, $l_\infty$-norm tomography~\cite{Kerenidis2019} provides strong guarantees on the worst-case error, though it demands quantum random access memory, which is an infeasible component in the near term. For noisy quantum devices with limited resources, efficient quantum state tomography is an urgent need.

\section{Methods} \label{sec3}

\subsection{Overview}
{When one performs large-scale fluid simulations on NISQ devices, the number or quality of qubits required by quantum fluid simulations may exceed the quantum computer's capability. Therefore, a key issue is how to simulate large-scale fluid on NISQ devices with a limited number of qubits. In classical algorithms, subspace methods can transform the solution of original linear systems into a series of lower-dimensional linear systems. Similarly, we can apply subspace methods to QLS to address scalability issues on a quantum computer. However, subspace methods require high-precision solutions for each linear system, without which the original linear system may fail to converge, imposing new challenges on QLS.}

\begin{figure}[ht]
    \centering
    \includegraphics[width=\linewidth]{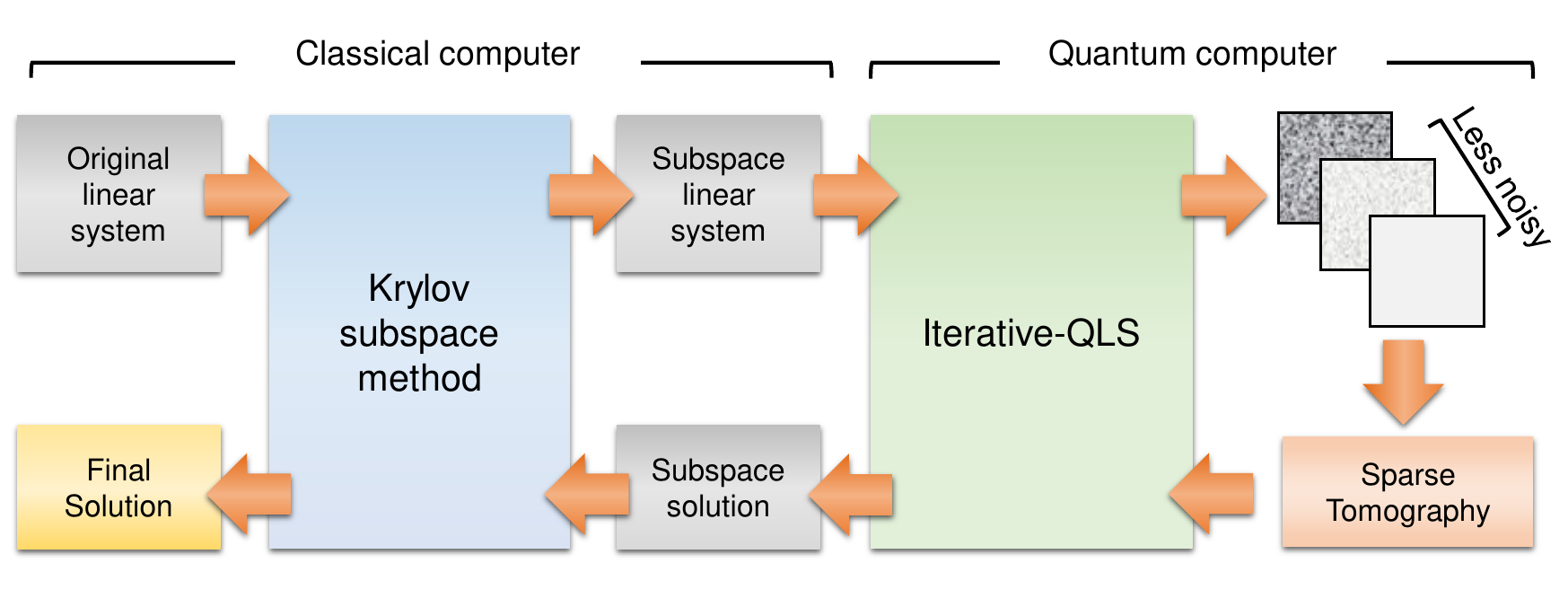}
    \caption{\textbf{Conceptual diagram of the proposed method.} In this paper, we focus on how to scale a noisy and limited-scale quantum computing resource to compute a large fluid simulation problem. This includes two stages: Iterative-QLS to improve the precision, and subspace method to achieve the scaling.}
    \label{fig:concept}
\end{figure}

To address these challenges, we implement a comprehensive approach to simulate a large-scale fluid flow with high precision on near-term quantum computers, as illustrated in Fig.~\ref{fig:concept}. 
First, we introduce an iterative approach for each small linear problem, called Iterative-QLS, to enable a high-precision quantum solver on noisy and limited-scale devices.
The cycle is repeated until the residual of the approximated solution is suppressed below an expected precision to meet the desired convergence condition demanded by the divide-and-conquer subspace method.
Second, the less-noisy while limited-scale solutions for different Krylov subspaces are synthesized to derive the large-scale solution.
To mitigate the quantum-classical information conversion bottleneck, a sparse tomography technique is developed to extract the desired information from the readout more efficiently.

\subsection{Iterative-QLS: Improving the precision of noisy quantum linear solver via iterative scheme}\label{sec:4}

\begin{figure}[ht]
\centering
\includegraphics[width = 0.75\textwidth]{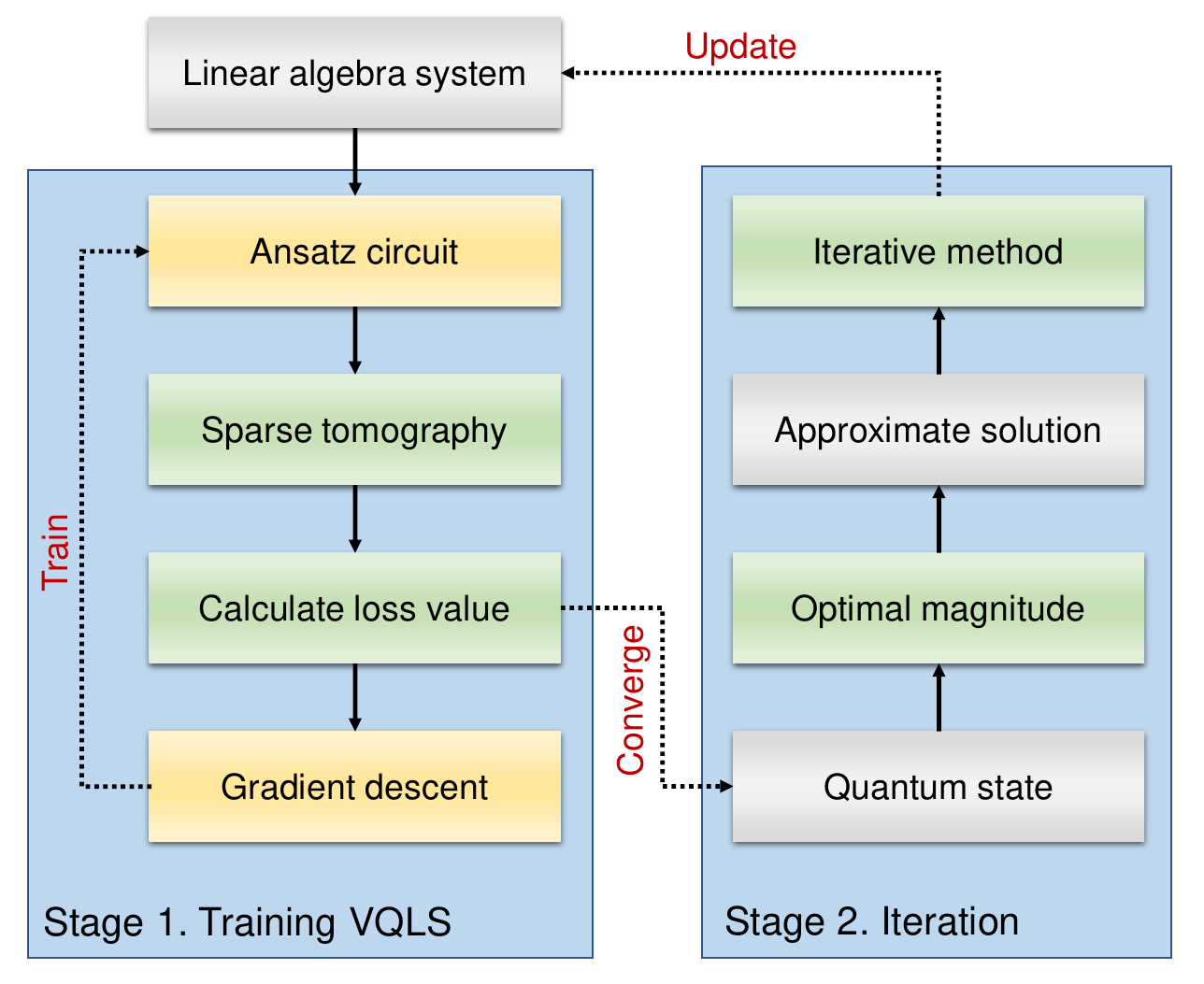}
\caption{\textbf{Flowchart of Iterative-QLS.}}
\label{fig:Iterative-QLS}
\end{figure}

To fulfill the requirements of performing large-scale flow simulation, improving the precision of a single VQLS becomes the first step. Here, we propose Iterative-QLS by introducing iterations beyond each VQLS, as shown in Fig.~\ref{fig:Iterative-QLS}. In each loop, we first derive an approximate solution $\tilde{\boldsymbol{x}}_i$ by executing a VQLS. We employ a variant of VQLS which is similar to a standard VQLS process, where we exclude the Hadamard-Test and replace it with a loss function based on the results of tomography. \ref{app:iqls-classicalcompute} shows how to evaluate the loss function and its gradient.

After executing VQLS, the tomography process should be executed to extract the data from the quantum state. In this paper, we propose an efficient method to extract a real-valued quantum state, named ``Sparse tomography'' which is introduced in the subsequent section~\ref{sec:sparse_tomo}.

Secondly, we evaluate the optimal magnitude based on the principle of $l_2$ norm.
To obtain the optimal magnitude, we propose the ``principle of minimum $l_2$ norm''. Assuming the noisy quantum state of the solution to the system of equations provided by VQLS is denoted as $|y\rangle$, the magnitude $L_y$ needs to be determined using certain optimal strategies. Here, $L_y$ should be the solution of the following optimization problem:
\begin{equation}
\underset{L_y}{\arg\min}~\Vert\boldsymbol{b}-L_yA|y\rangle\Vert_2.
\label{eq:trans}
\end{equation}
Let $\boldsymbol{z}=A|y\rangle$, with non-zero elements $z_i\ne 0$, $i\in[1,k]$. The solution to Eq.~\eqref{eq:trans} is explicitly given by
\begin{equation}
L_y=\left({\sum_{i=1}^{k}b_iz_i}\right)/\left({\sum_{i=1}^{k}z_i^2}\right).
\label{eq:min-2-norm}
\end{equation}
Then, the approximate solution to the system of equations is $\tilde{\boldsymbol{x}}=L_y|y\rangle$. 
Now the residual is defined to be $\boldsymbol{r}_i=\boldsymbol{b}-A\tilde{\boldsymbol{x}}_i$. 
By solving the new equation of residuals
$A\boldsymbol{y}=\boldsymbol{r}_i$, the approximate solution can be updated to $\tilde{\boldsymbol{x}}_{i+1}=\tilde{\boldsymbol{x}}_i+\boldsymbol{y}$. Notably, based on Eq.~\eqref{eq:trans}, we have 
\begin{equation}\label{eq:residual}
\Vert\boldsymbol{r}_{i+1}\Vert_2=\Vert\boldsymbol{r}_i-A\boldsymbol{y}\Vert_2\le\Vert\boldsymbol{r}_i\Vert_2.
\end{equation}
As shown in Eq.~\eqref{eq:residual}, the residual is strictly monotonically non-increasing and the equality holds only if $\langle b|A|y\rangle = 0$. 

To illustrate the idea of Iterative-QLS, we provide a short version in Alg.~\ref{alg:Iterative-QLS-simple}. A detailed algorithm procedure can be found in \ref{app:iqls-alg}, including the training of VQLS and iterations.

\begin{algorithm}[ht]
    \caption{Iterative-QLS (the detailed version in Appendix Alg.~\ref{alg:Iterative-QLS})}\label{alg:Iterative-QLS-simple}
    \begin{algorithmic}[1]
        \Require{Linear equations $A\boldsymbol{x}=\boldsymbol{b}$. Convergence criteria $\epsilon_c$. Convergence criteria of loss value $\epsilon_l$.}
        \State Give an initial solution $\tilde{\boldsymbol{x}}$.
        \State Obtain the residual $\boldsymbol{r}=\boldsymbol{b}-A\tilde{\boldsymbol{x}}$.
        \While{$\Vert \boldsymbol{r}\Vert_2>\epsilon_c$}
        \State Construct the Hamiltonian $M = A^{\dagger}\left( I-|\boldsymbol{r}\rangle\langle \boldsymbol{r}| \right)A$.
        \State Train \textbf{VQLS} process until satisfying the precision criteria $\epsilon_l$.
        \State Obtain the classical vector $\hat{\boldsymbol{y}}$ through \textbf{Sparse tomography}.
        \State Obtain the magnitude $L_y$ through \textbf{principle of minimum $l_2$ norm}.
        \State $\boldsymbol{y} \gets L_y\cdot \hat{\boldsymbol{y}} $
        \State Update $\tilde{\boldsymbol{x}}=\tilde{\boldsymbol{x}}+\boldsymbol{y}$.
        \State Obtain the residual $\boldsymbol{r}=\boldsymbol{b}-A\tilde{\boldsymbol{x}}$.
        \EndWhile
        \State \textbf{Return} $\tilde{\boldsymbol{x}}$.
    \end{algorithmic}
\end{algorithm}

\subsection{Scaling quantum linear solver to arbitrary scale via the subspace method}\label{sec:subspace}

To enhance scalability,
we develop the subspace method inspired by the classical renowned Krylov subspace method and the generalized minimal residual (GMRES) method \cite{SAAD2003}.
Instead of solving an $n$-dimensional linear system intractable large for near-term quantum computers,
we iteratively build small spaces of lower dimension $m \ll n$ and find the approximate solution therein.
More specifically, we consider the $m$-order Krylov subspace $\mathcal{K}_m$ spanned by basis $\left\{\boldsymbol{b}, A\boldsymbol{b}, A^2\boldsymbol{b}, ..., A^m \boldsymbol{b}\right\}$. This integration of Iterative-QLS and subspace method is implemented via a quantum-classical hybrid procedure.
As shown in Fig.~\ref{fig:concept}, the subspace is constructed on classical computers, and a subspace linear system is generated and solved by Iterative-QLS. The algorithm is presented in Alg.~\ref{alg:Sub-VQLS-simple}. 

\ref{app:sub-vqls} provides theoretical details of the subspace method. We conduct convergence analysis and show that the algorithm can converge with a very loose error bound, i.e. $\epsilon<1$ for a single system solved by Iterative-QLS. Also, we show that the approximate solution $\hat{x}\in\mathcal{K}_m$ turns out to be a good estimation of $x$ when $m$ is large enough and with a restart scheme, as detailed in \ref{app:sub-vqls-converg}.

Iterative-QLS is chosen as the internal solver for the subspace method primarily due to its high-precision features. Replacing Iterative-QLS with a single VQLS process may cause divergence in the subspace process, as each integration is not guaranteed to sufficiently shrink the residual, considering the error produced in a real quantum computer. Furthermore, Iterative-QLS can be replaced by a faster and more accurate quantum solver, which remains a subject for future works.

The size of the subspace can be determined by the capability of the internal QLS process. By providing a more powerful quantum computer, one can utilize a larger subspace size to accelerate the iteration process, an idea widely discussed in literature~\cite{joubert1994convergence,liesen2000computable,zitko2000generalization,simoncini2000convergence,liesen2004convergence,zou2023gmres}. Since QLS can solve $2^n$-dimensional linear system with $\operatorname{poly}(n)$ qubits and time complexity, it offers broad potential for quantum computing to accelerate a part of the fluid simulation procedure.

\begin{algorithm}[ht]
    \caption{Subspace method with Iterative-QLS (the detailed version in Appendix Alg.~\ref{alg:Sub-VQLS})}\label{alg:Sub-VQLS-simple}
    \begin{algorithmic}[1]
        \Require{Linear equations $A\boldsymbol{x}=\boldsymbol{b}$. Convergence criteria $\epsilon$. Subspace dimension $m$.}
        \State Give an initial solution $\tilde{\boldsymbol{x}}$.
        \State Obtain the residual $\boldsymbol{r} = \boldsymbol{b} - A \tilde{\boldsymbol{x}}$ and $\beta = \Vert \boldsymbol{r} \Vert_2$.
        \While{$\beta > \epsilon$}
        \State Construct subspace linear system $H \boldsymbol{y} = \beta\boldsymbol{e}_1$ with coefficient matrix $V$ by $m$-order \textbf{GMRES} method.
        \State Solve the subspace linear system through \textbf{Iterative-QLS}.
        \State Update $\tilde{\boldsymbol{x}} = \tilde{\boldsymbol{x}} + V\boldsymbol{y}$.
        \State Obtain the residual $\boldsymbol{r} = \boldsymbol{b} - A \tilde{\boldsymbol{x}}$ and $\beta = \Vert \boldsymbol{r} \Vert_2$.
        \EndWhile
        \State \textbf{Return} $\tilde{\boldsymbol{x}}$.
    \end{algorithmic}
\end{algorithm}

\subsection{Improving readout via sparse tomography}\label{sec:sparse_tomo}

The concept of sparse sampling has been adopted in many domains which require fast data conversion from quantum to classical. $l_\infty$-norm tomography was the first proposed in constructing a deep convolutional neural network~\cite{Kerenidis2019}. It offered an efficient method to generate such a sparse sample with $\mathcal{O}(\log N/\epsilon^2)$ time complexity, where the error of each entry is less than $\epsilon$. It was further applied to quantum algorithms~\cite{Chen2022, li2022quantum, xue2021quantum} that involve iterations where each iteration must convert quantum data to classical. In Ref.~\cite{Chen2022}, the application of sparse tomography to various CFD problems is studied, showing that sparse tomography, acting as an ``importance-sampling'' scheme, is available for the quantum-enhanced finite volume method.

However, $l_\infty$-tomography requires quantum random access memory (QRAM)~\cite{PhysRevLett.100.160501} to store the output and perform state preparation, as well as an additional qubit, resulting in extra costs for implementing the controlled version of the entire process. However, QRAM is widely believed to be unrealistic in the near term~\cite{jaques2023qram}, thus $l_\infty$-tomography is impracticable in near-term quantum computers. To address these limitations, a new method called ``sparse tomography'' is introduced in this study to achieve the same sparse sampling task without utilizing QRAM or any additional qubits.

The major idea of sparse tomography is the exploitation of the shadow tomography~\cite{Aaronson2018,Huang2020}, which is a novel tool to measure $O(\log N)$ times to compute $N$ expectation values by exploiting a classical postprocessing procedure named ``classical shadow''. In our algorithm, we construct the classical shadow by defining a computation graph to determine the sign of the quantum state. The algorithm's procedure is shown in Alg.~\ref{alg:STM} and the complexity and proof can be found in \ref{app:sparse-tomography}. Because the algorithm does not use QRAM and it only involves measurements on various measurement bases (refer to Ref.~\cite{Huang2020} for the concept about measurement bases), this procedure is efficient and completely capable for NISQ devices.

\begin{algorithm}[ht]
  \caption{Sparse tomography}\label{alg:STM}
  \begin{algorithmic}[1]
    \Require{A quantum algorithm that produces a $N$-dimensional real-valued state $|x\rangle$. Error criteria $\epsilon$}.
    \State Sample $|x\rangle$ with $36\log N/\epsilon^2$ copies to produce a sparse probability vector $\boldsymbol{p}$, and the position vector $\boldsymbol{v}$.
    \State Construct graph $\mathcal{G}=(V,E)$ with $\log\epsilon / \log v(1-p)$ random connected spanning subgraphs, where each subgraph contains all nodes with least edges ($N - 1$ edges). For each subgraph, prepare all sign-determination Hamiltonians $\{H_X^{i,j}\}$, which contains all $(i,j)$ pairs where $(i,j)\in E$.
    \State Use shadow tomography \cite{Aaronson2018,Huang2020} to calculate these Hamiltonians with $\delta$ error threshold, where $\delta = 1 / \sqrt{36\log N / \epsilon^2}$, and assign the graph's edge $E$ with the measurement outcomes. Use the outcomes' sign to assign the sign for each node.
    \State Collect the results for all subgraphs and assign the sign with the least violation to the measurement outcome. Finally, we obtain the sign $\boldsymbol{s}$ for this sparse vector.
    \State Return a classical vector $\boldsymbol{x}\gets \boldsymbol{s}\cdot\sqrt{\boldsymbol{p}}$. $\sqrt{\boldsymbol{p}}$ means a element-wise square-root of the vector $\boldsymbol{p}$.
  \end{algorithmic}
\end{algorithm}

\section{Experiments on a quantum computer} \label{sec:experiments}

\subsection{Experimental setup}
We mainly utilize VQLS as the base solver for building Iterative-QLS. To offer a comprehensive supplementary description of the computational process in this work, we provide a tutorial of VQLS in~\ref{app:intro_vqls}. In the VQLS process, the initial solution $\tilde{\boldsymbol{x}}$ is always chosen as $\boldsymbol{0}$. The initial ansatz parameter $\boldsymbol{\theta}_0$ is determined as $\boldsymbol{\theta}_0 ={\arg\min_{\theta_0^m}} L(\boldsymbol{\theta}_0^m)$, where $\boldsymbol{\theta}_0^m$ is generated by randomly sampling $m$ times from a uniform distribution over the interval $[0, 2\pi)$. 

The experiments are conducted on a superconducting quantum computer, named ``Wukong'' which is fabricated by Origin Quantum. The detailed specification of Wukong is shown in \ref{app:hardware}. No additional quantum error mitigation technique~\cite{cai2023quantum} is employed in the following experiments.

\begin{figure}[ht]
\centering
\includegraphics[width = .7\linewidth]{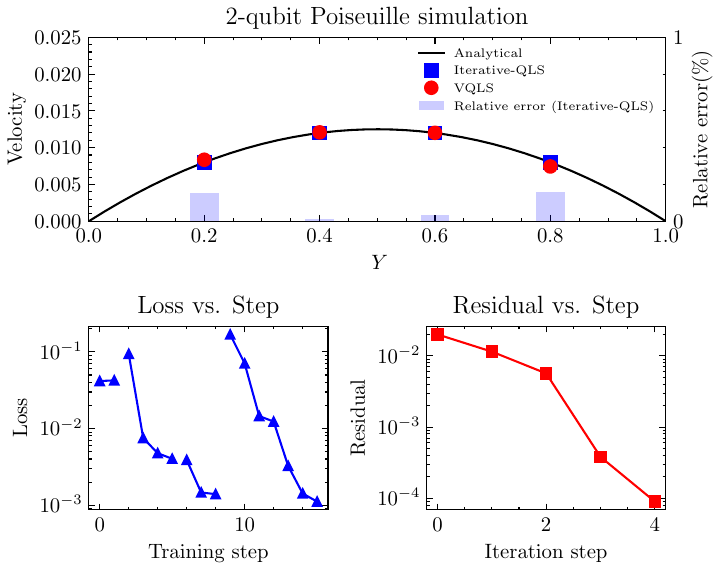}
\begin{tikzpicture}[overlay]  
  \node at (-9,7.3) {(a)};
  \node at (-9,3.5) {(b)};
  \node at (-4.2,3.5) {(c)};
  \node[scale=0.7] at (-8.18,2.25) {\textcolor{red}{1}};
  \node[scale=0.7] at (-7.6,2.3) {\textcolor{red}{2}};
  \node[scale=0.7] at (-6.88,1.2) {\textcolor{red}{3}};
  \node[scale=0.7] at (-6.1,2.25) {\textcolor{red}{4}};
  \node[scale=0.7] at (-3.7,2.9) {\textcolor{red}{init}};
  \node[scale=0.7] at (-3.15,2.6) {\textcolor{red}{1}};
  \node[scale=0.7] at (-2.4,2.3) {\textcolor{red}{2}};
  \node[scale=0.7] at (-1.7,1.2) {\textcolor{red}{3}};
  \node[scale=0.7] at (-1,1.1) {\textcolor{red}{4}};
\end{tikzpicture}
\caption{\textbf{Experimental results of Poiseuille flow.} (a) Quantitative comparison of velocity $u$ of 2D Poiseuille flow among results of Iterative-QLS solution $u_q$ (blue square) on four points with two qubits, single execution of VQLS (red dots), and analytical solutions $u_a$ (black line). Relative error for Iterative-QLS is also plotted, defined as $|u_q/u_a-1|\times10^2$. (b) Loss values change with increasing iteration steps. Each line segment is an independent VQLS training process, corresponding to an iteration step of Iterative-QLS. (c) Residual change with increasing iteration steps. The number marks the iteration step count, which shows the correspondence between (b) and (c).}
\label{fig:p-flow}
\end{figure}
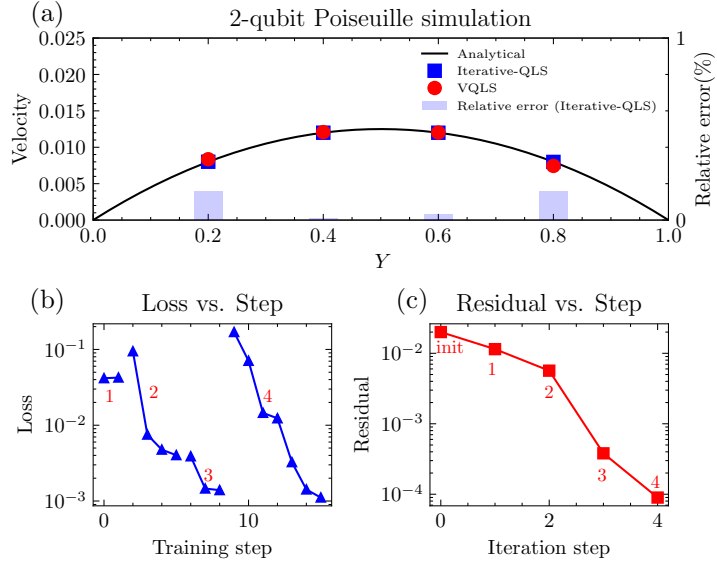 

\subsection{Numerical simulation of Poiseuille flow}\label{sec:Poiseuille}
The governing equation of 2D incompressible Poiseuille flow is
\begin{equation}
\frac{\partial u}{\partial t}=-\frac{1}{\rho}p_x+\mu\frac{\partial^2 u}{\partial y^2},
\label{eq:gve}
\end{equation}
where $x,y$ represent the coordinates in the flow direction and transverse direction, $u$ is the velocity in the x-direction, $\rho$ is the fluid density, $p_x$ is the pressure gradient in the x-direction, $\mu$ is the fluid viscosity. The boundary condition is $y=\pm h: u=0$, where $h$ represents the distance between the wall and the center line of the channel. The analytical solution to Eq.~\eqref{eq:gve} is $u_a = {p_x}\left( h^2-y^2 \right)/{2\rho\mu}$. 

Considering a uniform Cartesian grid for discretization in the y-direction, an implicit scheme to discretize the time term, and a central difference scheme to discretize the diffusion term, the discretized governing equation can be constructed as follows
\begin{equation}
\frac{u^+_j-u_j}{\Delta t}=\mu\frac{u_{j-1}^+-2u_j^++u_{j+1}^+}{\Delta y^2}-\frac{p_x}{\rho},
\label{eq:dgve}
\end{equation}
where the superscript $+$ means the velocity in the next time step which is unknown, the subscript $j-1$, $j$, and $j+1$ represent three adjacent grid points in the y-direction, $\Delta t$ is the discrete time step, and $\Delta y$ represents the grid size in the y-direction. Combining the discretized governing equations on all grid points in the y-direction, we obtain a system of linear equations
\begin{equation} 
A\boldsymbol{x}^+=\boldsymbol{x}-p_x\Delta t/\rho.
\label{eq:le}
\end{equation}
For a steady flow, it should hold that $\boldsymbol{x}^+=\boldsymbol{x}$, thus iteratively solving linear system Eq.~\eqref{eq:le} is equivalent to directly solving linear system as follows:
\begin{equation}
A\boldsymbol{x}=\boldsymbol{b},\quad A=\begin{bmatrix}
\beta &\alpha&0     &0 \\
\alpha&\beta &\ddots&0 \\
0     &\ddots&\ddots&\alpha \\
0     &0     &\alpha&\beta \\
\end{bmatrix}, \quad\boldsymbol{b} = \gamma\begin{bmatrix}
1 \\
1 \\
\vdots \\
1 \\
\end{bmatrix},
\end{equation}
where $\alpha = -\mu\Delta t/\Delta y^2, \beta = -2\alpha, \gamma = -p_x\Delta t/\rho$. The matrix of the linear system is always a static tridiagonal matrix.

In this work, we set $p_x = -0.1$, $\mu = 1$, $\rho = 1$, and $\Delta t = 0.01$. The number of grids in the y-direction, denoted as $N_y$, is determined by the number of qubits used but always satisfies $N_y\Delta y=1$. For Iterative-QLS, we set the loss function tolerance $\epsilon_l=10^{-3}$, convergence tolerance $\epsilon_c = 10^{-4}$ and the number of shots for each circuit is $10^4$.

A two-qubit experiment is conducted as a preliminary step to validate the feasibility of Iterative-QLS. From the results shown in Fig.~\ref{fig:p-flow}(a), it can be seen that Iterative-QLS accurately solved the problem of 2D incompressible Poiseuille flow. The quantum results closely matched the analytical solutions with a relative error of less than 0.2\%. We also plot the results produced by a single execution of VQLS. By comparing the results of Iterative-QLS and a single execution of VQLS, we conclude that Iterative-QLS can effectively suppress the error generated in VQLS.

Fig.~\ref{fig:p-flow}(b) illustrates the variation of the loss value during the training process. Each line segment represents an independent VQLS training process. Fig.~\ref{fig:p-flow}(c) shows the change of loss during the iteration and Iterative-QLS reaches the convergence criteria after 4 iterations, with the residual decreasing to below $10^{-4}$. Here, the number marked in subfigure (b) corresponds to the iteration step count in subfigure (c). It can be seen that the residual value exhibits a clear decreasing trend, which confirms the successful execution of Iterative-QLS. 

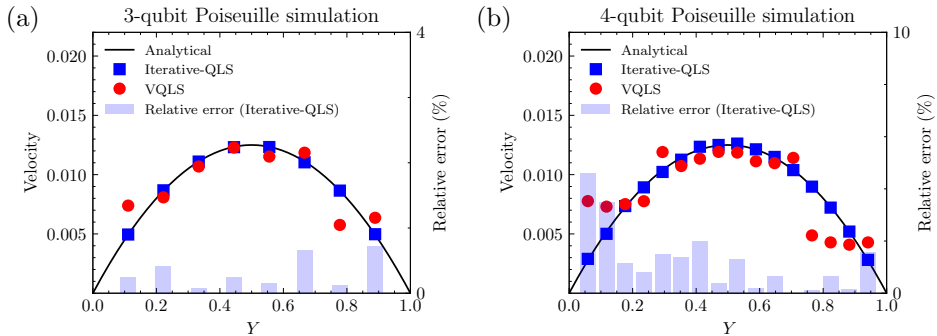
\begin{figure}[ht]
    \centering
    \includegraphics[width = .9\linewidth]{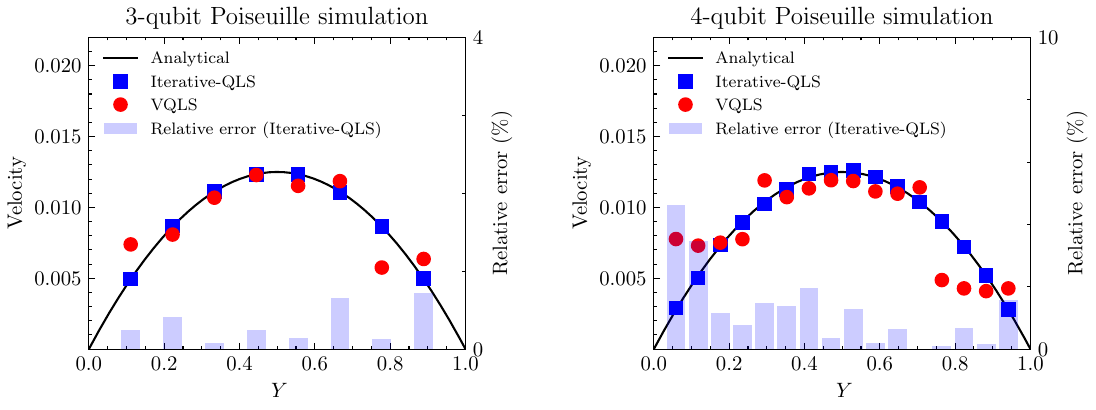}    
    \begin{tikzpicture}[overlay]  
      \node at (-12.2,4.3) {(a)};
      \node at (-6,4.3) {(b)};
    \end{tikzpicture}
    \caption{\textbf{Experimental results of Poiseuille flow with more qubits.} (a) Experimental results for 3-qubit Poiseuille simulation on 8 points. (b) Experimental results for 4-qubit Poiseuille simulation on 16 points. We plot results of Iterative-QLS solution $u_q$ (blue square) on four points with two qubits, single execution of VQLS (red dots), and analytical solutions $u_a$ (black line). Relative error for Iterative-QLS is also plotted, defined as $|u_q/u_a-1|\times10^2$.}
    \label{fig:p-flow-34}
\end{figure}

\begin{figure}[ht]
    \centering
    \includegraphics[width = .9\linewidth]{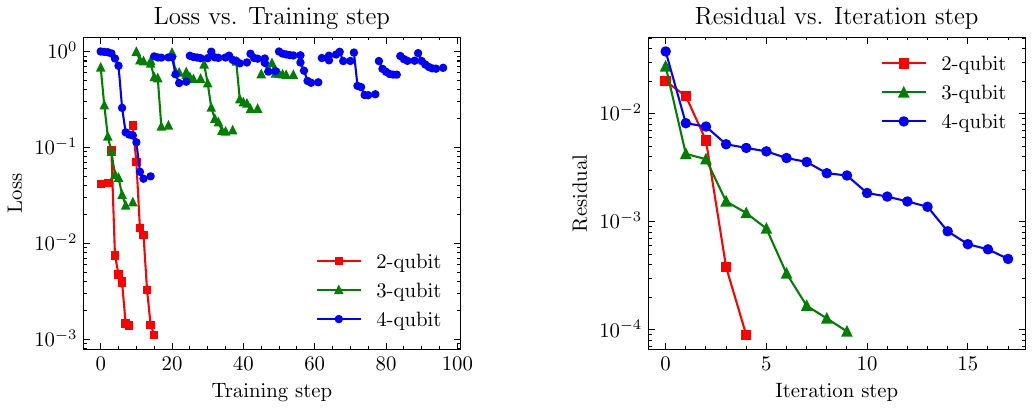}    
    \begin{tikzpicture}[overlay]  
      \node at (-12.2,4.3) {(a)};
      \node at (-5.5,4.3) {(b)};
    \end{tikzpicture}
    \caption{\textbf{Convergence history of Poiseuille flow experiments. } (a) Loss values change with increasing iteration steps. One can observe that the average loss increases with a growing number of qubits. (b) Residual change with increasing iteration steps. Note that two-qubit results shown in Fig.~\ref{fig:p-flow-34} are also plotted here as a comparison.}
    \label{fig:p-flow-34-converg}
\end{figure}
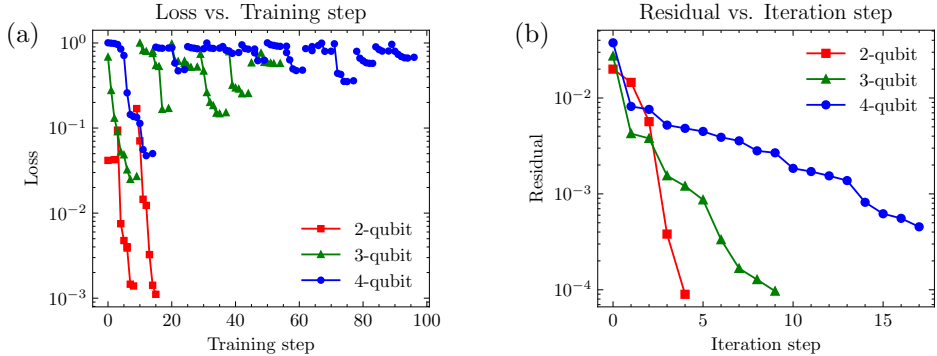

{We also demonstrate 3-qubit and 4-qubit experiments, as shown in Fig.~\ref{fig:p-flow-34}. The results also show qualitative consistency between the analytical solution and the experimental results. {Compared to a single execution of VQLS, Iterative-QLS can indeed output results with higher precision. We observe that the performance decreases with an increasing number of qubits. To understand this phenomenon, we compare the convergence history from 2-qubit to 4-qubit in Fig.~\ref{fig:p-flow-34-converg}. The training procedure of VQLS, as demonstrated in subfigure~(a), displays the original solution quality produced by each single VQLS, which clearly shows that more qubits will result in a decrease in the performance of each VQLS. Although Iterative-QLS manages to suppress the error of all cases, as shown in subfigure~(b), the increasing error from single VQLS finally leads to a decreasing performance of the convergence.}

{Iterative-QLS utilizes iterations to suppress the error generated in VQLS.} A similar counterpart is quantum error mitigation (QEM)~\cite{cai2023quantum}, which also utilizes redundant circuits to achieve lower error on NISQ devices. Compared to QEM, our technique focuses more specifically on the linear solver problem, resulting in a much lower overhead. The number of circuits executed in two-qubit, three-qubit, and four-qubit experiments is less than 20, 60, and 100, respectively. QEM techniques like probabilistic error cancellation~\cite{van2023probabilistic} or zero-noise extrapolation~\cite{kim2023evidence} should first introduce a heavy-cost error calibration circuit, then utilize thousands of circuit instances to obtain one result. Therefore, our technique is much more efficient than QEM in our experiments. Moreover, QEM only reduces circuit-level error, while our method can further reduce the inherent trainability issue of VQLS~\cite{wang2021noise}. Finally, QEM does not conflict with our method, which could be an effective way to scale Iterative-QLS when the size of the quantum circuit grows larger in the future.

{Last but not least, Iterative-QLS is not limited to the current version of VQLS and our implementation. Iterative-QLS could produce more accurate results by designing an ansatz with better expressibility~\cite{PRXQuantum.3.010313, Nakaji_2021}, or other types of quantum solvers~\cite{jaksch2023variational, liu2024variational, demirdjian2022variational, PhysRevA.101.010301}, along with a stronger quantum computer in the future.}

\subsection{Acoustic wave propagation simulation}\label{sec:Acoustic}
In this section, linear acoustic wave propagation is simulated on NISQ devices. The acoustic wave propagation can be described with the linearized Euler equations as follows:
\begin{equation} \label{eqn:leuler-1}
	\begin{aligned} 
		\frac{\partial p}{\partial t} &= - \bar{\rho} c^2 (\frac{\partial u}{\partial x} + \frac{\partial v}{\partial y}) - \bar{u} \frac{\partial p}{\partial x} + S(t,x,y),\\
		\frac{\partial u}{\partial t} &= - \frac{1}{\bar{\rho}} \frac{\partial p}{\partial x} - \bar{u} \frac{\partial u}{\partial x}-\bar{v} \frac{\partial u}{\partial y}, \\
		\frac{\partial v}{\partial t} &= - \frac{1}{\bar{\rho}} \frac{\partial p}{\partial y} - \bar{u} \frac{\partial v}{\partial x}- \bar{v} \frac{\partial v}{\partial y}.
	\end{aligned}
\end{equation}
The source term has the following form:
\begin{align} \label{eqn:sound_source}
	S(t,x,y) = \epsilon \ \exp ^{-\alpha \frac{(x-x_{0})^2 +(y-y_{0})^2}{b^{2}}}\sin (\omega t),
\end{align}
where angle velocity $\omega$ and frequency \textit{f} is related in $\omega = 2 \pi f$. $(x_0,y_0)$ is the location of the sound source and $\epsilon$ is the source amplitude. 

To accurately capture the sound waves, numerical schemes with low dissipation and low dispersion are required to discretize the equations~\cite{zhang2023wcs}. The dispersion relation preserving (DRP) scheme \cite{tam1993dispersion} is one of the most efficient schemes widely applied in computational aeroacoustics (CAA). The seven-point DRP scheme \cite{tam1995computational} is applied to spatial discretization in this work. The derivative is discretized with a first-order Euler implicit scheme for time advancement to achieve higher advanced efficiency and more robust numerical stability. Furthermore, to avoid non-physical wave reflection near boundaries, the non-reflecting radiation boundary condition is imposed on all boundaries \cite{tam1996radiation}.

Applying the numerical methods above, the governing equations are discretized into the following form:
\begin{align}
    &\begin{aligned}\label{eqn:leuler-d1}
    p^{n}_{i,j}=p^{n+1}_{i,j} + \bar{\rho} c^2 \frac{\Delta t}{\Delta x} \sum_{m=0}^{M} a_{m} u^{n+1}_{i+m,j}  
		  + \bar{\rho} c^2 \frac{\Delta t}{\Delta y} \sum_{m=0}^{M} a_{m} v^{n+1}_{i,j+m}+\bar{u} \frac{\Delta t}{\Delta x} \sum_{m=0}^{M} a_{m} p^{n+1}_{i+m,j},\\
    \end{aligned}\\
    &\begin{aligned}\label{eqn:leuler-d2}
    u_{i,j}^{n} =u_{i,j}^{n+1}+\frac{1}{\bar{\rho}}\frac{\Delta t}{\Delta x}\sum_{m=0}^{M}a_{m}p_{i+m,j}^{n+1}+\bar{u}\frac{\Delta t}{\Delta x}\sum_{m=0}^{M}a_{m}u_{i+m,j}^{n+1} +\bar{v}\frac{\Delta t}{\Delta y}\sum_{m=0}^{M}a_{m}u_{i,j+m}^{n+1},
    \end{aligned}\\
    &\begin{aligned}\label{eqn:leuler-d3}
    v_{i,j}^{n} =v_{i,j}^{n+1}+\frac{1}{\bar{\rho}}\frac{\Delta t}{\Delta y}\sum_{m=0}^{M}a_{m}p_{i+m,j}^{n+1}+\bar{u}\frac{\Delta t}{\Delta x}\sum_{m=0}^{M}a_{m}v_{i+m,j}^{n+1} +\bar{v}\frac{\Delta t}{\Delta y}\sum_{m=0}^{M}a_{m}v_{i,j+m}^{n+1}.
    \end{aligned}
\end{align}

As a result, a large sparse linear system of equations is obtained, of which the dimension is $3N$, where $N$ is the number of grid points. In practice, the grid point number is relatively large, so the linear system is too large for the current NISQ device to handle. Thus the subspace method can be used to reduce the problem size to match the capability of NISQ devices.

In this experiment, a 1000Hz sound source is placed at the center of a $[-1,1]\times[-1,1]$ square zone. The zone is uniformly meshed with $35 \times 35$ grid points. In addition, three layers of ghost points are extended outside the boundaries for better numerical treatment near boundaries, resulting in actual $41 \times 41$ grid points. To save time, the flow is advanced to a developed state with classical methods to get the initial flow field. The classical linear solver uses the LU decomposition method to yield results accurately enough. Starting with the initial flow field, the flow is solved with the subspace method on the quantum computer, and the wave propagates forward about $0.93$ wavelengths. The final pressure field is compared with the classical counterpart to validate the reliability.

\begin{figure}[H]
\centering
\includegraphics[width=.9\linewidth]{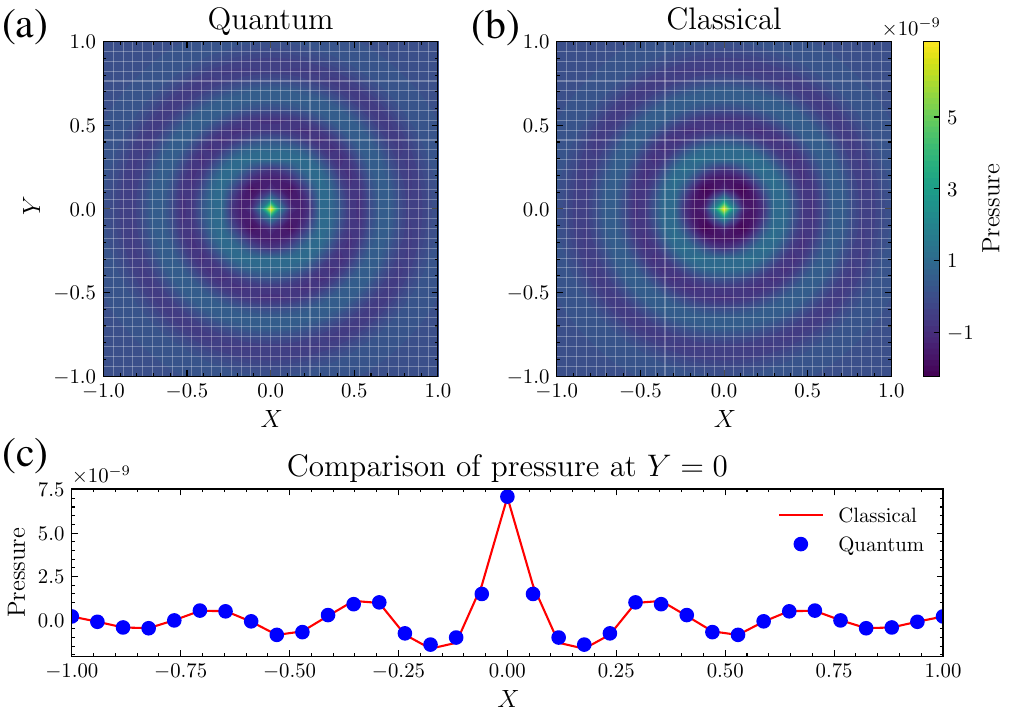}
 \caption{\textbf{Results of acoustic wave propagation.} (a) Pressure fluctuation contour computed with classical and quantum computers. (b) Quantitative comparison of pressure fluctuation of 2D acoustic wave propagation between results of Iterative-QLS solution (blue dots) with 2 qubits and classical solutions (red line).}
 \label{fig:caa_field_q}
\end{figure}

Starting with the same initial flow field, the flow is advanced with time step size. Fig.~\ref{fig:caa_field_q}(a) and (b) show the pressure fluctuation contour of the quantum and classical results respectively. Both results are in qualitative agreement. For quantitative comparison, the results of slices at $Y=0.0$ are shown in Fig.~\ref{fig:caa_field_q}(c). After a short evolution, a shift of acoustic waves propagating outward is observed, and the quantum result agrees with the classical result at phase and magnitude. Therefore, the proposed quantum approach is suitable for unsteady acoustic wave propagation and the simulation is reliable.

To explore the convergence property, we show the convergence history of the subspace method, including the external iteration by the subspace method and internal iteration by Iterative-QLS method. Fig.~\ref{fig:caa-2}(a) shows a part of the convergence history of the external iterations during simulation. For each time step, the residual converges after 3 subspace iterations, showing the good convergence property of the subspace method. To further investigate the convergence performance of internal iterations, a piece of convergence history of the iteration of Iterative-QLS during simulation is shown in Fig.~\ref{fig:caa-2}(b). It is observed that, in each subspace iteration of every time step, the residual of Iterative-QLS drops quickly during iteration, showing the efficient convergence of solving the linear system of equations with the subspace method. 

\begin{figure}[H]
\centering
\includegraphics[width = .9\linewidth]{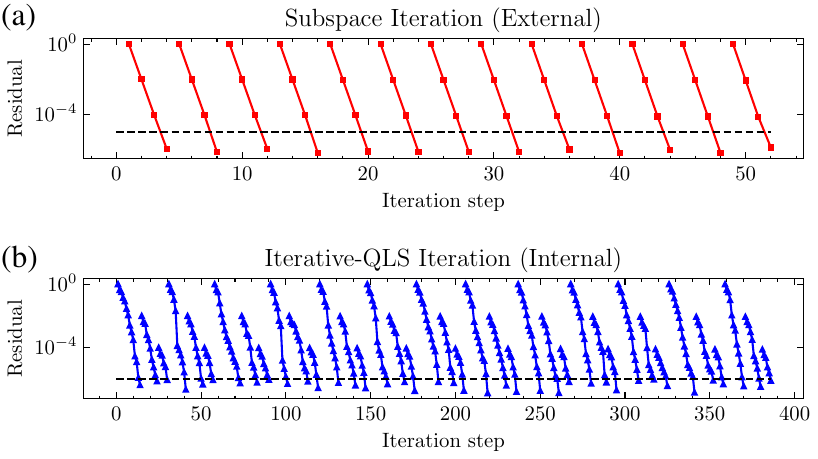}
\caption{\textbf{Convergence history details during the experiment. } (a) Convergence history of the subspace method during simulation. The dashed line is the convergence criteria ($10^{-5}$) of the subspace method. (b) Convergence history of Iterative-QLS to achieve high order during simulation. The dashed line is the convergence criteria ($10^{-6}$) of iteration of VQLS. }
\label{fig:caa-2}
\end{figure} 

\section{Conclusion and Outlook}\label{sec:7}

In this paper, we propose a novel approach to CFD leveraging the capabilities of near-term quantum computers, along with several enhancements to existing methods to enable effective CFD simulations. First, we introduce a high-precision quantum linear solver suitable for NISQ devices, called Iterative-QLS. This solver effectively mitigates errors that typically arise in quantum computing processes. Our simulation of a 2D Poiseuille flow on a near-term quantum computer demonstrates a relative error of less than $0.2\%$. To address the challenges in quantum state tomography, we develop a sparse tomography algorithm, which removes the dependency on quantum random access memory inherent in $l_\infty$-norm tomography. Second, we introduce the subspace method to scale the Iterative-QLS method for larger CFD problems, enhancing solution convergence and resulting in a quantum-classical hybrid approach. By employing the subspace method, we successfully simulate unsteady acoustic wave propagation on a $41 \times 41$ grid using a superconducting quantum computer, marking a significant advancement toward scalable quantum-enhanced CFD solutions.

{We highlight the importance of utilizing a hybrid scheme. If the entire $41 \times 41$ grid is simulated entirely on a quantum computer, one should employ at least 13 high-quality qubits. However, in our experiments, two noisy qubits are sufficient to produce high-precision and large-scale results. Our method significantly reduces the cost of simulation on a quantum computer to fewer qubits and effectively suppresses the noise. Therefore, our proposed method provides a feasible way to scale the existing method to large-scale fluid simulation with near-term quantum computers, especially those with limited qubit numbers and gate performance. }

{The last question is whether our method is a scalable method to achieve potential quantum speedup. The subspace method, as a quantum-classical hybrid approach, allocates the subspace problem to quantum computers, leaving the remaining part computed on the classical method. By numerical examples in~\ref{app:scalability}, we show that the number of subspace iteration steps decreases significantly by investing every one more qubit. Then, one can expect quantum computers can provide speedup in the subspace problem at a certain amount of qubits, which will be left as future work.}

Identifying suitable application scenarios for current near-term quantum computers is crucial for advancing the field of CFD. The algorithms presented in this paper serve as a bridge between quantum and classical computing, enabling a range of practical CFD problems to be executed on near-term quantum computers with varying qubit capacities. Furthermore, our proposed method can be extended to other potential applications in computational science, paving the way for the practical utilization of near-term quantum computers in CFD and beyond.

\section{Conflict of interest}\label{sec:9}
This paper declares no conflict of interest.

\section{Acknowledgments}\label{sec:10}
This work is supported by the National Key Research and Development Program of China (Grant No. 2023YFB4502500) and the Aeronautical Science Foundation of China (Grant No. 2022Z073004001). We also acknowledge the financial support provided by the China Academy of Aerospace Aerodynamics (CAAA).

\appendix

\section{Quantum hardware specification}\label{app:hardware}
All experiments are implemented based on the OriginQ quantum cloud. The processor employed to conduct the experiment is the ``OriginQ-Wukong'' superconducting quantum processor, whose average $T_1$ time is \SI{14.51}{\micro\second}, average $T_2$ time is \SI{1.84}{\micro\second}, average fidelity of single-qubit gate is 0.9965, average fidelity of CZ gate is 0.9686, and the number of qubits is 72.

\begin{figure}[ht]
\centering
\includegraphics[width = 0.35\textwidth]{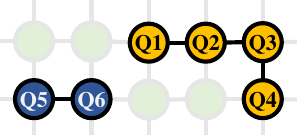}
\caption{The chosen qubits used in numerical experiments of 2D incompressible Poiseuille flow (yellow) and linear acoustic wave propagation (blue).}
\label{fig:topology}
\end{figure} 

Fig.~\ref{fig:topology} shows the chip-topology of Wukong, and the chosen qubits used in experiments of this work. The parameters of used qubits are shown in Tab.~\ref{tab:qubits}. The fidelity of CZ gates is shown in Tab.~\ref{tab:cz_gate_fidelity}.

\begin{table}[H]
  \centering
  \caption{Summary of parameters of qubits}
  \label{tab:qubits}
  \begin{tabularx}{\textwidth}{c*{6}{>{\centering\arraybackslash}X}}
  \toprule
  \textbf{}              &\textbf{Q1}&\textbf{Q2}&\textbf{Q3}&\textbf{Q4}&\textbf{Q5}&\textbf{Q6}\\
  \midrule
  $T_1$ $(\mu s)$        &6.21       &35.47      &17.27      &11.08      &11.37      &7.18     \\
  $T_2$ $(\mu s)$        &1.82       &1.21       &1.47       &1.45       &2.84       &1.21     \\
  $F_{00}$               &0.9750     &0.9572     &0.9562     &0.9328     &0.892      &0.9242    \\
  $F_{11}$               &0.8508     &0.8546     &0.9518     &0.8428     &0.8528     &0.8922     \\
  Single gate fidelity   &0.9973     &0.9976     &0.9972     &0.9978     &0.9982     &0.9979     \\
  \bottomrule
  \end{tabularx}
\end{table}

\begin{table}[H]
  \centering
  \caption{CZ gate fidelity between qubits}
  \label{tab:cz_gate_fidelity}
  \begin{tabularx}{\textwidth}{c*{4}{>{\centering\arraybackslash}X}}
  \toprule
  \textbf{Qubit pairs}        &\textbf{Q1-Q2}&\textbf{Q2-Q3}&\textbf{Q3-Q4}&\textbf{Q5-Q6}\\
  \midrule
  CZ gate fidelity       &0.9583        &0.9822       &0.9821       &0.9807      \\
  \bottomrule
  \end{tabularx}
\end{table}

\section{Theory details of Iterative-QLS}\label{app:iqls}

In this section, we provide theory details of Iterative-QLS.

\subsection{Full algorithm of the Iterative-QLS}\label{app:iqls-alg}

\begin{algorithm}[H]
  \caption{Iterative-QLS}\label{alg:Iterative-QLS}
  \begin{algorithmic}[1]
    \Require{Linear equations $A\boldsymbol{x}=\boldsymbol{b}, \boldsymbol{b}\ne\boldsymbol{0}$. Difference step size $\Delta$. Convergence criteria  $\epsilon_c$. Convergence criteria of loss value $\epsilon_l$.Learning rate $\lambda$}.
    \State Give a initial solution $\tilde{\boldsymbol{x}}$.
    \State Obtain the residual $\boldsymbol{r}=\boldsymbol{b}-A\tilde{\boldsymbol{x}}$ and the Hamiltonian $M = A^{\dagger}\left( I-|r\rangle\langle r| \right)A$.
    \If{$\Vert r\Vert_2>\epsilon_c$}
        \State Goto line 8.
    \Else
        \State Goto line 27.
    \EndIf
    \State Set $k = 0$ and give a initial ansatz parameter $\boldsymbol{\theta}_k$.
    \State Run hardware-efficient ansatz circuits $U_{\theta_k}$ and get the quantum state $|\psi_k\rangle=U_{\theta_k}|0\rangle$ by the \textbf{sparse tomography} method.
    \State Obtain the loss value $L_{k} = \langle\psi_k|M|\psi_k\rangle$
    \While{$L_{k} > \epsilon_l$}
        \State Prepare $\boldsymbol{\theta}_k^{*} = \boldsymbol{\theta}_k + \Delta$.
        \State Run hardware-efficient ansatz circuits $U_{\theta_k^*}$ and obtain $|\psi_k^{*}\rangle=U_{\theta_k^*}|0\rangle$.
        \State Calculate $L_{k}^{*} = \langle\psi_k^{*}|M|\psi_k^{*}\rangle$.
        \State Obtain the gradient $\boldsymbol{g}_k$ using parameter-shift rule.
        \State Update $\boldsymbol{\theta}_{k+1} = \boldsymbol{\theta}_{k} - \lambda\boldsymbol{g}_k$.
        \State Run hardware-efficient ansatz circuits $U_{\theta_{k+1}}$ and obtain $|\psi_{k+1}\rangle=U_{\theta_{k+1}}|0\rangle$.
        \State Calculate $L_{k+1}=\langle\psi_{k+1}|M|\psi_{k+1}\rangle$.
        \If{$L_{k} < L_{k+1}$}
            \State \textbf{break}
        \EndIf
        \State Set $k = k+1$.
    \EndWhile
    \State Set $|y\rangle = |\psi_k\rangle$ and obtain $\boldsymbol{y}$ based on the \textbf{principle of minimum $L_2$ norm}.
    \State Update $\tilde{\boldsymbol{x}}=\tilde{\boldsymbol{x}}+\boldsymbol{y}$.
    \State \textbf{Goto} line 2.
    \State \textbf{Return} $\tilde{\boldsymbol{x}}$.
  \end{algorithmic}
\end{algorithm}

\subsection{Introduction of VQLS}\label{app:intro_vqls}

In our work, we primarily use the Variational Quantum Linear Solver (VQLS). Below, we provide a detailed introduction to the execution process of VQLS.

For an $N$-dimensional system of linear equations $A\boldsymbol{x}=\boldsymbol{b}$, where $\Vert \boldsymbol{b}\Vert_2\ne0$, if there exists a non-zero solution $\boldsymbol{x}=L_x|x\rangle$ with $L_x=\Vert\boldsymbol{x}\Vert_2$ being the magnitude, it can be easily verified that quantum state $|x\rangle$ is the ground state of the Hamiltonian $H=A^{\dagger}\left( I-|b\rangle\langle b| \right)A$, satisfying
\begin{equation}
A^{\dagger}\left( I-|b\rangle\langle b| \right)A|x\rangle=0.
\end{equation}
Moreover, a loss function can be defined as $L=\langle\psi|H|\psi\rangle$, which characterizes the ``distance'' between any quantum state $|\psi\rangle$ and the ground state $|x\rangle$. By achieving a sufficiently small loss value $L$, the corresponding quantum state $|\psi\rangle$ can be considered a good approximation of $|x\rangle$.

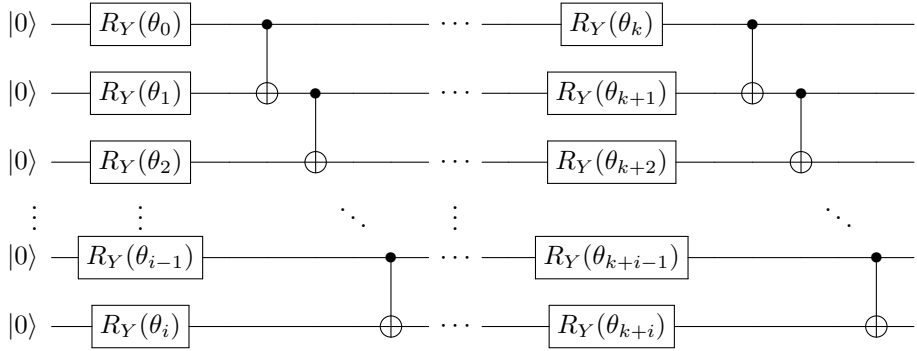
\begin{figure}[ht]
\centerline
{
\Qcircuit @C=1.0em @R=1.0em
{
&\lstick{\ket{0}}&\gate{R_Y(\theta_0)}    &\qw&\ctrl{1}&\qw     &\qw   &\qw     &\qw&\cdots&\qquad\quad&\qw&\gate{R_Y(\theta_k)}      &\qw&\ctrl{1}&\qw     &\qw   &\qw     &\qw \\
&\lstick{\ket{0}}&\gate{R_Y(\theta_1)}    &\qw&\targ   &\ctrl{1}&\qw   &\qw     &\qw&\cdots&\qquad\quad&\qw&\gate{R_Y(\theta_{k+1})}    &\qw&\targ   &\ctrl{1}&\qw   &\qw     &\qw \\
&\lstick{\ket{0}}&\gate{R_Y(\theta_2)}    &\qw&\qw     &\targ   &\qw   &\qw     &\qw&\cdots&\qquad\quad&\qw&\gate{R_Y(\theta_{k+2})}    &\qw&\qw     &\targ   &\qw   &\qw     &\qw \\
&\lstick{\vdots} &\vdots                  &   &        &        &\ddots&        &   &\vdots&           &   &                          &   &        &        &\ddots&        &    \\
&\lstick{\ket{0}}&\gate{R_Y(\theta_{i-1})}&\qw&\qw     &\qw     &\qw   &\ctrl{1}&\qw&\cdots&\qquad\quad&\qw&\gate{R_Y(\theta_{k+i-1})}&\qw&\qw     &\qw     &\qw   &\ctrl{1}&\qw \\
&\lstick{\ket{0}}&\gate{R_Y(\theta_i)}    &\qw&\qw     &\qw     &\qw   &\targ   &\qw&\cdots&\qquad\quad&\qw&\gate{R_Y(\theta_{k+i})}    &\qw&\qw     &\qw     &\qw   &\targ   &\qw \\
}
}
\caption{An example of hardware-efficient ansatz.}
\label{fig:ansatz}
\end{figure}

In VQLS, the quantum state is constructed by the ansatz circuit $U_\theta$ where $|\psi\rangle = U_\theta |0\rangle$, implying that the search for approximate solutions is confined within the ``ansatz space''. Fig.~\ref{fig:ansatz} illustrates the hardware-efficient ansatz circuit, which is also the circuit used in this work. After introducing the ansatz, the problem to be solved is transformed into the following optimization problem
\begin{equation}
\underset{\boldsymbol{\theta}}{\arg\min}~L\left(\boldsymbol{\theta}\right),
\end{equation}
where $\boldsymbol{\theta}$ represents the parameters of the ansatz circuits.

This optimization problem is typically solved by the gradient descent method \cite{RUDER2016}
\begin{equation}
\boldsymbol{\theta}^{k+1} = \boldsymbol{\theta}^k - \lambda \boldsymbol{g}.
\end{equation}
Here, $\lambda$ is the learning rate, and $\boldsymbol{g} = \partial L / \partial \boldsymbol{\theta}$ is the gradient of the loss function. There are multiple ways to obtain the gradient. For example, one can directly observe every component of $\boldsymbol{g}$ \cite{XU2021} using the Hadamard-Test circuit. If a small enough loss value has been found, the information of the corresponding quantum state can be extracted through quantum tomography in the end.

\subsection{Classical computation for loss value and its gradient}\label{app:iqls-classicalcompute}

The original VQLS algorithm employs distinct circuit configurations to evaluate the loss value and discern individual gradient components. Traditionally, deriving these scalar quantities has relied on implementing Hadamard-Test circuits. However, executing a universal Hadamard-Test circuit on contemporary NISQ devices poses significant challenges, primarily due to their intricate nature.

The sparse tomography method yields an approximate sparse vector $\hat{\boldsymbol{x}}$ of the quantum state $|x\rangle$. Subsequently, the value of the loss function can be efficiently computed as $L \approx \hat{\boldsymbol{x}}^TH\hat{\boldsymbol{x}}$ on classical computers, thereby circumventing the need for complex Hadamard-Test circuits.

To address the complexities of directly observing gradient components, our study utilizes numerical approximations instead of exact determinations. Specifically, by employing the parameter-shift rule, the gradient's components can be estimated as follows:
\begin{equation}
g_i = \frac{L(\theta_i+\frac{\pi}{2}) - L(\theta_i-\frac{\pi}{2})}{2}.
\label{eq:grad}
\end{equation}

\section{Theory details of subspace method} \label{app:sub-vqls}

In this section, we provide theory details of subspace methods. In \ref{app:sub-vqls-fullalg}, we provide the full algorithm of the subspace method integrated with Iterative-QLS. Then the convergence analysis is conducted to show the noise resilience of the subspace method, in \ref{app:sub-vqls-converg}. 

\subsection{Full algorithm of the subspace method with Iterative-QLS}\label{app:sub-vqls-fullalg}
\begin{algorithm}[H]
    \caption{Subspace method with Iterative-QLS}\label{alg:Sub-VQLS}
    \begin{algorithmic}[1]
        \Require{Linear equations $A\boldsymbol{x}=\boldsymbol{b}$. Convergence criteria $\epsilon$. Subspace dimension $m$.}
        \State Give an initial solution $\tilde{\boldsymbol{x}}$.
        \State Obtain the residual $\boldsymbol{r} = \boldsymbol{b} - A \tilde{\boldsymbol{x}}$ and $\beta = \Vert \boldsymbol{r} \Vert_2$.
        \If{ $\beta \le \epsilon $}
        \State \textbf{Goto} line 37.
        \EndIf
        \State $\boldsymbol{v}_1 = \boldsymbol{r}_0 / \beta$, $\boldsymbol{\xi} = \beta \boldsymbol{e}_1$.
        \For{ $j \in [1, m]$}
        \State $\boldsymbol{w}_j = A \boldsymbol{v}_j$.
        \For{ $i \in [1,j]$}
        \State $h_{ij} = (\boldsymbol{w}_j, \boldsymbol{v}_i)$.
        \State $\boldsymbol{w}_j = \boldsymbol{w}_j - h_{ij}\boldsymbol{v}_i$.
        \EndFor
        \State $h_{j+1,j} = \Vert \boldsymbol{w}_{j} \Vert_2$.
        \For{ $i \in [1, j-1]$}
        \State $\begin{bmatrix}h_{i,j}\\h_{i+1,j}\end{bmatrix}=\begin{bmatrix}c_i&s_i\\-s_i&c_i\end{bmatrix}\begin{bmatrix}h_{i,j}\\h_{i+1,j}\end{bmatrix}$.
        \EndFor
        \If{ $h_{j+1,j} = 0$}
        \State $m = j$.
        \State \textbf{break}
        \EndIf
        \State $\boldsymbol{v}_{j+1}=\boldsymbol{w}_j/h_{j+1,j}$.
        \If{ $|h_{j,j}| > |h_{j+1,j}|$}
        \State $c_j = 1/\sqrt{1+\tau^2}, s_j=c_j\tau, \tau = h_{j+1,j}/h_{j,j}$.
        \Else
        \State $s_j = 1/\sqrt{1+\tau^2}, c_j=s_j\tau, \tau = h_{j,j}/h_{j+1,j}$.
        \EndIf
        \State $h_{j,j} = c_jh_{j,j}+s_jh_{j+1,j}$.
        \State $h_{j+1,j} = 0$.
        \State $\begin{bmatrix}\xi_j\\ \xi_{j+1}\end{bmatrix}=\begin{bmatrix}c_j&s_j\\-s_j&c_j\end{bmatrix}\begin{bmatrix}\xi_j\\0\end{bmatrix}$.
        \If{ $|\xi_{j+1}| < \beta\epsilon$}
        \State $m = j$.
        \State \textbf{break}
        \EndIf
        \EndFor
        \State Solve the subspace linear system $H \boldsymbol{y} = \boldsymbol{\xi}$ through Iterative-QLS. The matrix $H$ is composed of elements $h$.
        \State Update $\tilde{\boldsymbol{x}} = \tilde{\boldsymbol{x}} + V\boldsymbol{y}$, where $V = [\boldsymbol{v}_1, \cdots , \boldsymbol{v}_j]$.
        \State \textbf{Goto} line 1.
        \State \textbf{Return} $\tilde{\boldsymbol{x}}$.
    \end{algorithmic}
\end{algorithm}

\subsection{Convergence analysis}\label{app:sub-vqls-converg}
We start with the well-known conclusion for GMRES that there must exist a solution from $n$-th order subspace.
\begin{thm}[Maximum order of the subspace from the dimension]    
    We have $n$ the dimension of the matrix and the exact solution $\boldsymbol{x}$, the Krylov subspace method is guaranteed to have produced a solution $\hat{\boldsymbol{x}}$ such that ${\lVert \hat{\boldsymbol{x}}-\boldsymbol{x}\rVert}<\epsilon{\lVert \boldsymbol{x}\rVert}$, i.e. $\hat{\boldsymbol{x}} \in \mathcal{K}_{n}$. 
\end{thm}

Here, $\mathcal{K}_n$ represents an $n$-th order subspace of the linear equation $A\boldsymbol{x}=\boldsymbol{b}$, that is,
\begin{equation}
    \mathcal{K}_n = \left\{\boldsymbol{b}, A\boldsymbol{b}, A^2\boldsymbol{b}, ..., A^n \boldsymbol{b}\right\}.
\end{equation}
Then, we can show that when the condition number is given, the order of the subspace is limited to the condition number $\kappa$.
\begin{thm}[Maximum order of the subspace from the condition number]
    When selecting $m \in \mathcal{O}(\kappa)$, given the matrix's condition number $\kappa(A)=\kappa$, $\lVert A\rVert<1$ and the exact solution $\boldsymbol{x}$, the Krylov subspace method is guaranteed to have produced a solution $\hat{\boldsymbol{x}}$ such that ${\lVert \hat{\boldsymbol{x}}-\boldsymbol{x}\rVert}<\epsilon{\lVert \boldsymbol{x}\rVert}$, i.e. $\hat{\boldsymbol{x}} \in \mathcal{K}_{\mathcal{O}(\kappa\log 1/\epsilon)}$. 
\end{thm}

\begin{pf}
    We can set $\hat{\boldsymbol{x}}\in \mathcal{O}_m = p(A)\boldsymbol{b}$, where $p$ is a polynomial of the maximum order of $m-1$, that is $p(x)=\sum_{i=0}^{m-1}a_i x^i$. We have
    \begin{equation}
        \frac{1-(1-x^2)^b}{x} = 4\sum_{j=0}^{b}(-1)^j\left[\frac{\sum_{i=j+1}^b C_{2b}^{b+i}}{2^{2b}} \right]\mathcal{T}_{2j-1}(x),
    \end{equation}
    where $\mathcal{T}_n(x)$ is the $n$-th order of the Chebyshev polynomial of the first kind. Truncating $b$ to the order $\widetilde{\mathcal{O}}(\kappa^2)$ is sufficient to approximate $\frac{1}{x}$ by $\frac{1-(1-x^2)^b}{x}$ in the subspace $[-1,-1/\kappa]\cup [1/\kappa, 1]$. Here we use $\widetilde{\mathcal{O}}$ to ignore the sublinear term, which does not affect the analysis.

    Now we approximate on the right-hand side by truncating the series to the order of $j_0\in \widetilde{\mathcal{O}}(\sqrt{b})$. Notice that 
    \begin{equation}\label{eq:truncation}
        \begin{aligned}
            &\left|4\sum_{j=j_0+1}^{b}(-1)^j\left[\frac{\sum_{i=j+1}^b C_{2b}^{b+i}}{2^{2b}} \right]\mathcal{T}_{2j-1}(x)\right|\\
            \leq &\left|\sum_{j=j_0+1}^{b}(-1)^j\sum_{i=j+1}^b e^{-j^2/b}\mathcal{T}_{2j-1}(x)\right|\leq 4be^{-j_0^2/b}.
        \end{aligned}        
    \end{equation}
    Therefore, $j_0$ has the order of $j_0\in \widetilde{\mathcal{O}}(\sqrt{b})$ to allow the truncation with a bounded error. Finally, we see that there must exist a polynomial $p(x) = \sum_{i=0}^{j_0}a_i\mathcal{T}_i(x)$ with at most $i$-th order where $j_0\in \mathcal{O}(\kappa\log 1/\epsilon)$. With $\boldsymbol{x}=A^{-1}\boldsymbol{b}$ and $\tilde{\boldsymbol{x}}=p(A)\boldsymbol{b}$, we can conclude that $\lVert \hat{\boldsymbol{x}}-\boldsymbol{x}\rVert\leq \epsilon\lVert \boldsymbol{x}\rVert$.
\end{pf}

This is a reverse application of the theorem in Ref.~\cite{Childs2017}. This theorem provides a strong guarantee that every linear equation can be approximated by a subspace with the same order of $\kappa$. The upper bound of the subspace's order can be tightened to $\mathcal{O}(\max (\kappa, n))$. Naturally, this does not mean we should initially select $\kappa$ as the subspace order, including two reasons. First, computing $\kappa$ requires time. Second, $\kappa$ is often at the order of $n$. Still, we are glad to see that the error decays exponentially by the truncation order $j_0$, from which we have the following result.

\begin{thm}[Error bound produced by subspace method]\label{Lemma:subspace_error}
    There exists a solution from the Krylov subspace with exponential decay of error, i.e. $\lVert \boldsymbol{x}-\hat{\boldsymbol{x}}\rVert \leq \mathcal{O}(e^{-(m/\kappa)^2})$ where $\hat{\boldsymbol{x}}\in \mathcal{K}_m$. 
\end{thm}

\begin{pf}
From Eq.~\eqref{eq:truncation}, we see that there exists a polynomial of $m$-th order $p_m(x)$ that suffices $|1/x-p_m(x)|<4be^{-m^2/b}$. Note that we only consider the decay rate from a growing $m$, so that $|1/x-p_m(x)|\leq \mathcal{O}(e^{-(m/\kappa)^2})$ where $b$ is truncated to $\widetilde{\mathcal{O}}(\kappa^2)$.
\end{pf}

One of the most common cases for applying the subspace method is to use a constant subspace's order $m$ and the restarting scheme. It keeps updating the residual with a more precise solution and finally obtains the solution. From this idea, we show the convergence of the subspace method with two steps: first, the convergence can be satisfied with an existing solution from the subspace $\mathcal{K}_m$; second, the convergence can be satisfied with a fixed residual produced by VQLS method.

\begin{thm}[Convergence condition from the Chebyshev polynomial method]\label{Lemma:chebyshev_err}
    There exists a subspace solution $\hat{\boldsymbol{x}}\in \mathcal{K}_m$ with a fixed $m$ and restarting scheme that satisfied any bounded error $\lVert \boldsymbol{x}-\hat{\boldsymbol{x}}\rVert \leq \epsilon$.
\end{thm}

\begin{pf}
    The starting residual of the equation is $\boldsymbol{r}_0=\boldsymbol{b}-A\boldsymbol{x}_0$. The equation will be converted to solve $A\boldsymbol{x}_1=\boldsymbol{r}_0$ and update the solution to $\boldsymbol{x}=\boldsymbol{x}_0+\boldsymbol{x}_1$. Therefore, this lemma is equivalent to showing that we can always provide a better solution within the $m$-th order Krylov subspace compared to $\boldsymbol{0}$ as the initial guess for the converted equation. Note that $\lVert \boldsymbol{x}-\hat{\boldsymbol{x}}\rVert \leq \mathcal{O}(e^{-(m/\kappa)^2})$, the Chebyshev polynomial of the first kind has already provided such a solution with better precision, given any $m$ and $b$.
\end{pf}

Next, we can naturally obtain the convergence condition for the subspace method.

\begin{thm}[Convergence of the integration of QLS and subspace method]\label{thm:convergence}
    The subspace method produces a solution with $\epsilon$ error as long as QLS finds a solution better than $\boldsymbol{0}$ as the initial guess. That is, we can solve the linear system with precision $\max(\mathcal{O}(e^{-m^2/\kappa^2}), \epsilon_q) < 1$ such that $C = \langle \psi|A^\dagger (I-|b\rangle\langle b|) A|\psi\rangle \leq \epsilon_q$. The number of iterations is bounded by 
    \begin{equation}
        \min\left(\frac{\log(\epsilon)}{\log(\epsilon_q)}, \kappa\right),
    \end{equation}
    where $\kappa$ is the condition number of the original linear system.
\end{thm}

\begin{pf}
    The convergence of the subspace method is satisfied for either of the following cases. The first case is that when QLS shrinks the residual by $\epsilon_q$ in each iteration. As the initial guess of the original equation can be considered as a constant factor, the number of iterations of the subspace method is bounded by $\log(1/\epsilon_q)$. The second case is a fixed bound produced by the Chebyshev polynomial. From Thm.~\ref{Lemma:subspace_error} and Thm.~\ref{Lemma:chebyshev_err}, there always exists an $\epsilon\in \mathcal{O}(e^{-(m/\kappa)^2})$ approximate solution when the order of the subspace is chosen to be $\mathcal{O}(\kappa)$. By the restarting scheme, this provides an iteration count of $\mathcal{O}(\kappa)$.
\end{pf}

It is obvious to see when $\kappa$ is larger, we are required to have a better $\epsilon_q$ (or extra iterations to the subspace method). We should note that this is only a lower bound for the convergence of the QLS. That is, by satisfying $\epsilon_q < 1$, one can always guarantee the convergence of the solution. In other words, Thm.~\ref{thm:convergence} provides an insight into our method that the QLS is not always required to compute an exact solution. Also, in the realistic test, the convergence can be faster. 

\section{Theory details of sparse tomography} \label{app:sparse-tomography}

In this section, we will provide rigorous proof of the sparse tomography given in the main text. Firstly, we show the definition of quantum sparse sampling.

\begin{problem} [Quantum sparse sampling]
For a $N$-dimensional real-valued quantum state $|x\rangle=\sum_{i=0}^{N-1}x_i|i\rangle$, we produce a classical vector $\hat{\boldsymbol{x}}$ that satisfies $\Vert \boldsymbol{x} - \hat{\boldsymbol{x}}\Vert_\infty <\epsilon$ with a number of copies of the quantum state. The classical vector has $k$ nonzero entries such that $k\ll N$. Here, real-valued means that every $x_i$ is a real number.
\label{rmk:sparse sampling}
\end{problem}

Then we refer to the existing approach for the sparse sampling problem, shown in Thm.~\ref{thm:linf_tomo} with its algorithm description in Alg.~\ref{alg:linf_tomo}.

\begin{thm}[$l_\infty$-norm tomography \cite{Kerenidis2019}]\label{thm:linf_tomo}
    Given access of unitary $U$ such that $U|0\rangle=|x\rangle$ and its controlled version, there exists an algorithm that produces an $l_\infty$ sampling for a real-valued quantum state with $\mathcal{O}(\log N/\epsilon^2)$ query complexity. The algorithm requires a constant number of queries to QRAM and an extra qubit. 
\end{thm}

\begin{algorithm}[H]
  \caption{$l_\infty$-norm tomography \cite{Kerenidis2019}}\label{alg:linf_tomo}
  \begin{algorithmic}[1]
	\Require{Error $\epsilon$, access to unitary $U: |0\rangle \to |x\rangle = \sum_{i\in[N]}x_i|i\rangle$, the controlled version of $U$, QRAM access.}
    \Ensure{Classical vector $\tilde{X}\in \mathbb{R}^N$, such that $\lVert\tilde{X}\rVert = 1$ and $\lVert\tilde{X}-\boldsymbol{x}\rVert_{\infty}<\epsilon$.}
	\State Measure $M = 36\log N/\epsilon^2$ copies of $|x\rangle$ in the standard basis and count $m_i$, the number of times the outcome $i$ is observed. Store $\sqrt{p_i} = \sqrt{m_i/M}$ in QRAM data structure.
    \State Create $M = 36\log N/\epsilon^2$ copies of the state $\frac{1}{\sqrt{2}}|0\rangle\sum_{i\in[N]}x_i|i\rangle + \frac{1}{\sqrt{2}}|1\rangle\sum_{i\in[N]}\sqrt{p_i}|i\rangle$.
    \State Apply a Hadamard gate on the first qubit to obtain
    \begin{equation*}
    |\phi\rangle = \frac{1}{2}\sum_{i\in[N]}\left( \left(x_i+\sqrt{p_i}\right)|0,i\rangle + \left(x_i-\sqrt{p_i}\right)|1,i\rangle \right)
    \end{equation*}
    \State Measure both registers of each copy in the standard basis, and count $n(0, i)$ the number of times the outcome $(0, i)$ is observed.
    \State Set $\sigma(i) = +1$ if $m(0,i) > 0.4Mp_i$ and $\sigma(i) = -1$ otherwise.
    \State Output the unit vector $\tilde{X}$ such that $\forall i\in [M], \tilde{X}_i = \sigma_i\sqrt{p_i}$.
  \end{algorithmic}
\end{algorithm}

Based on the analysis in Ref.~\cite{Kerenidis2019}, the original problem is converted to a sign-determination problem where we have to correctly determine the sign of the sparse-sampled terms. The result of this section is given in Thm.~\ref{thm:sparse_tomo}.

\begin{thm}[Efficient sparse tomography without QRAM and extra qubits]\label{thm:sparse_tomo}
There exists an algorithm that produces a sparse sampling for an $N$-dimensional real-valued quantum state with $\epsilon$ error with $\mathcal{O}(\log N/\epsilon^2)$ time complexity. The algorithm can run without using QRAM and any extra qubit, i.e. all procedures are only based on measurements on different bases.
\label{thm:sparse-tomography}
\end{thm}

Similar to $l_\infty$ tomography, the first step is to sample $36\log N/\epsilon^2$ copies of the quantum state to produce a positive valued vector to the quantum state. Note that $k<36\log N/\epsilon^2$ is always satisfied so that the classical vector is sparse when $36\log N/\epsilon^2<N$. Now the major target is to determine the relative phase of these sparse entries of the quantum state. The technique is to apply shadow tomography to compute the sign of each term. 

\begin{thm}[Sign determination between two computational bases]
Given two computational bases $|i\rangle$ and $|j\rangle$ in a real-valued quantum state $|x\rangle$, there exists a Hamiltonian $H_X^{i,j}$ that determines whether signs of $|i\rangle$ and $|j\rangle$ are different. The sign is determined by $s = \sgn(\langle H_X^{i,j}\rangle)$, and the number of measurements is $\mathcal{O}(\log N/\epsilon^2)$ to obtain the correct sign with a constant successful rate. Here, $H_X^{i,j}$ is defined as follows:
\begin{equation}
H_X^{i,j} = \bigotimes_{u=0}^{n}H_u
\end{equation}
where $H_u = X$ if $u$-th digit is different in $|i\rangle$ and $|j\rangle$, otherwise $H_u = |0\rangle\langle0|$. We can also define $H_Z$ with a similar definition where $H_u = Z$ if the digit is different. Note that the expectation value of $(I+H_Z)/2$ is easy to calculate by using the previous sampling results.
\label{thm:sign-deter}
\end{thm}

\begin{pf}
Obviously, we show that $H_X^{i,j}$ measures the probability of $(\alpha+\beta)/2$ and $(\alpha-\beta)/2$, where $\alpha = \langle i|x\rangle$ and $\beta=\langle j|x\rangle$. If the signs of $\alpha$ and $\beta$ are the same, we have an accurate $\langle H_X^{i,j}\rangle >0$. If they are different, then $\langle H_X^{i,j}\rangle <0$. To determine the sign of $\langle H_X^{i,j}\rangle$, we have to perform measurements to obtain an approximation $O$ of $\langle H_X^{i,j}\rangle$ that separates $|\alpha+\beta|$ and $|\alpha - \beta|$, which can be written as

\begin{equation}
\langle H_X^{i,j}\rangle=\frac{1}{2}(|\alpha+\beta|-|\alpha - \beta|)=\min(|\alpha|, |\beta|)\ge\frac{1}{\sqrt{36\log N/\epsilon^2}}
\end{equation}

For the number of measurements $M$, and defining $\delta = 1/\sqrt{36\log N/\epsilon^2}$, we use the Hoeffding's inequality and obtain
\begin{equation}
P(|O-\langle H_X^{i,j}\rangle|>\delta)\le 2\exp(-M\delta^2/2)
\end{equation}
Provided $M=12/\delta^2$, we have $P(|O-\langle H_X^{i,j}\rangle|>\delta)<2e^{-3}<0.005$. Here $|O-\langle H_X^{i,j}\rangle|>\delta$ denotes that the sign is measured correctly.
\end{pf}

Note that the complexity is mainly induced by the least probability of the measurement so that we can improve the Thm.~\ref{thm:sign-deter} to a more compact form by providing the probabilities from these two computational bases. 

\begin{thm}
(Sign determination between two computational bases with given probabilities). For two computational bases $|i\rangle$ and $|j\rangle$ in a real-valued quantum state $|x\rangle$, there exists a Hamiltonian $H_{X}^{i,j}$ that determines whether signs of $|i\rangle$ and $|j\rangle$ are different. The sign is determined by $s=\operatorname{sgn}(\langle H_{X}^{i,j}\rangle)$, and the number of measurements is $\mathcal{O}[\frac{1}{\min(\sqrt{p_i}, \sqrt{p_j})}]$ to obtain the correct sign with a constant successful rate. Here $p_i$ and $p_j$ are the probabilities of $i$ and $j$ computational bases extracted from step 1 of Alg.~\ref{alg:STM}. 
\end{thm}

Now we consider the scenario where all signs must be determined. If we generate a sparse sample with $p$ entries, we can create a fully connected graph $\mathcal{G}=(V, E)$ with $v$ nodes and $v(v-1)/2$ edges. Here, each node represents a computational basis, and each edge represents the measurement value of $H_{X}^{i,j}$. As the number of measurements becomes sufficiently large, each edge will be assigned a value of $+1$ or $-1$ when the two connected nodes have the same or different signs, respectively. 

Given that each edge is successfully measured with a constant success rate $p$, we can traverse this graph randomly in $v$ steps and determine all signs successfully with a probability of $p^v$. Repeating this random traversal $k$ times will result in a success rate of approximately $1-\left(1-p^v\right)^k \sim 1-\left(v\left(1-p\right)\right)^k$. Therefore, we can choose $k=(\log \epsilon)/\left(\log {v\left(1-p\right)}\right)$ to ensure the algorithm succeeds with a probability of $1-\epsilon$, with the number of measurements being $\mathcal{O}(Mvk)$, where $M$ still represents the number of measurements for each Hamiltonian, and $\mathcal{O}(vk)$ denotes the number of Hamiltonians that need to be computed.

While $p$ is generated from Hoeffding's inequality with a number of measurements $M$, where $p\geq 1-2\exp(-M\delta^2/2)$, we can derive an explicit sample complexity for this algorithm, given by
\begin{equation}
\begin{aligned}
    \mathcal{O}(Mvk)&=\mathcal{O}\left(\frac{18 \log^2 N\cdot(\log(36) + \log(\log N) - 2\log \epsilon)}{\epsilon^2 (\log(\log N) + 12)}\right)\\
    &=\widetilde{\mathcal{O}}\left(\frac{\log^2 N\cdot\left|\log \epsilon\right|}{\epsilon^2}\right).
\end{aligned}
\end{equation}
This result implies that the algorithm should sample $\widetilde{\mathcal{O}}({\epsilon^{-2}}{\log^2 N})$ times to generate all required routes in this fully connected graph.

Now we optimize this algorithm with shadow tomography. Since all the sign-determination Hamiltonians can be computed beforehand, we can efficiently sample these Hamiltonians using shadow tomography with a protocol requiring only $\mathcal{O}(\log vk/\delta^2)$ measurements. Therefore, the number of measurements can be further optimized to
\begin{equation}
    \begin{aligned}        
    \mathcal{O}(M \log(vk)) &= \mathcal{O}\left(\frac{\log N \cdot \log\left(\frac{\log N}{\epsilon^2}\right)}{\epsilon^2}\right)\\
    &=\widetilde{\mathcal{O}}\left(\frac{\log N}{\epsilon^2}\right).
    \end{aligned}
\end{equation}

This complexity aligns with the complexity in the first step of the algorithm. Therefore, the sparse-tomography method can be executed with an explicit algorithm, with time complexity also being $\mathcal{O}(\log N/\epsilon^2)$.

\section{Examples for the discussion about scalability of the subspace method}\label{app:scalability}

{Here, we demonstrate examples to show the scalability of the subspace method. The integration of the subspace method on the flow simulation is promising when more qubits are provided to enable a larger subspace size $m=2^n$ where $m$ is the size of the subspace, and $n$ corresponds to the number of qubits utilized in the internal QLS. In Fig.~\ref{fig:sub-vqls-subspace}, we solve $5043$-dimensional tridiagonal matrices, the same size as the acoustic wave propagation problem presented in the main text, with $m$ varies from $2^4$ to $2^{10}$. We test two matrices with different condition numbers and plot the number of subspace steps after reaching the same convergence criteria. In the context of the integration of the subspace method with a QLS, the number of steps corresponds to the number of internal QLS executions on a quantum computer.}

{The results show an exponential decay in subspace steps with respect to the number of qubits, which represents that the subspace method could run faster if sending a larger subspace problem into the quantum solver. Therefore, in this example, the subspace method has shown great scalability: one more qubit investment would effectively get an exponential reward in terms of speed, showing the potential ability for quantum speedup.}

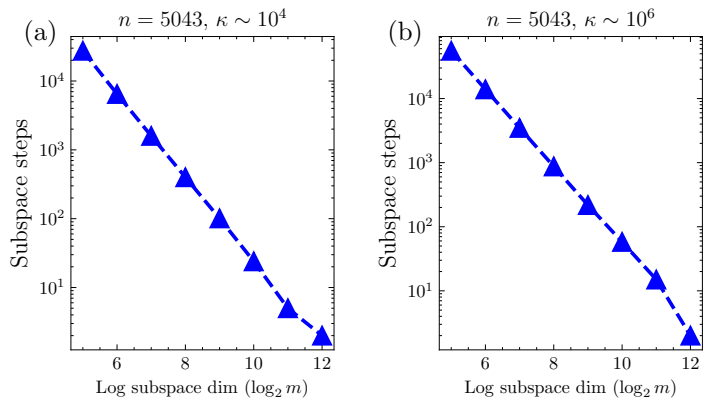
\begin{figure}[ht]
    \centering
    \includegraphics[width=.7\linewidth]{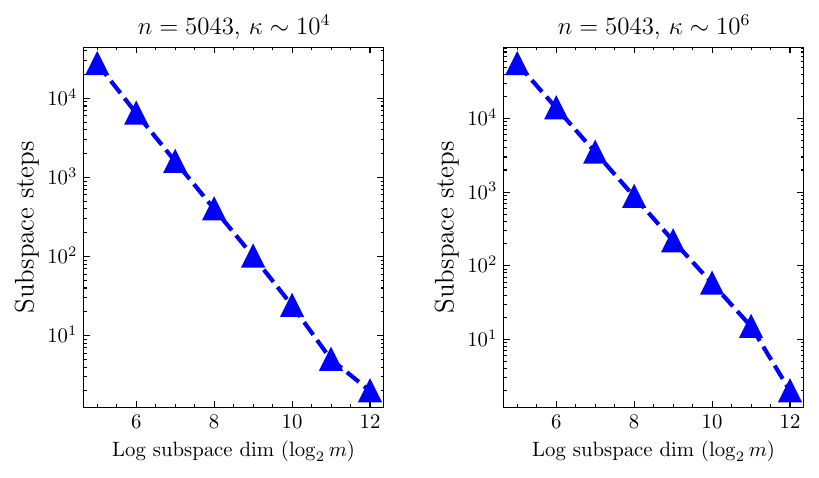}
    \begin{tikzpicture}[overlay]  
        \node at (-9,5) {(a)};
        \node at (-4.2,5) {(b)};
    \end{tikzpicture}
    \caption{\textbf{Results of solving 5043-dimensional matrices with different subspace sizes.} (a) Subspace steps with different subspace dimensions. The condition number is $\kappa\sim 10^4$. (b) Subspace steps with different subspace dimensions. The condition number is $\kappa\sim 10^6$.}
    \label{fig:sub-vqls-subspace}
\end{figure}

% \section{Declaration of generative AI and AI-assisted technologies in the writing process}

% During the preparation of this work, the authors used ChatGPT in order to polish the language. After using this tool/service, the authors reviewed and edited the content as needed and take full responsibility for the content of the publication.

\bibliographystyle{elsarticle-num} 
\bibliography{reference}

\end{document}